\documentclass[12pt]{article}
\pdfoutput=1

\usepackage{graphicx}
\usepackage{epstopdf}
\usepackage{amsmath}
\usepackage{amsfonts}
\usepackage{amssymb}
\usepackage{color}

\newdimen\tableauside\tableauside=1.0ex
\newdimen\tableaurule\tableaurule=0.4pt
\def\e{\mathrm{e}}

\def\R{\zeta}
\def\Rtwo{\langle\zeta^2\rangle}
\def\ep{\epsilon}
\def\beq{\begin{eqnarray}}
\def\eeq{\end{eqnarray}}
\def\epsosc{\epsilon_{\rm osc}}
\def\lsim{\mathrel{\rlap{\lower3pt\hbox{\hskip0pt$\sim$}}
     \raise1pt\hbox{$<$}}}         %less than or approx. symbol
\def\gsim{\mathrel{\rlap{\lower3pt\hbox{\hskip0pt$\sim$}}
     \raise1pt\hbox{$>$}}}         %less than or approx. symbol

\usepackage{subfigure}
\usepackage{amsmath}
\usepackage{amsfonts}
\usepackage{amssymb}
\usepackage{calligra}

\newdimen\tableauside\tableauside=1.0ex
\newdimen\tableaurule\tableaurule=0.4pt
\newdimen\tableaustep
\def\phantomhrule#1{\hbox{\vbox to0pt{\hrule height\tableaurule width#1\vss}}}
\def\phantomvrule#1{\vbox{\hbox to0pt{\vrule width\tableaurule height#1\hss}}}
\def\sqr{\vbox{%
  \phantomhrule\tableaustep
  \hbox{\phantomvrule\tableaustep\kern\tableaustep\phantomvrule\tableaustep}%
  \hbox{\vbox{\phantomhrule\tableauside}\kern-\tableaurule}}}
\def\squares#1{\hbox{\count0=#1\noindent\loop\sqr
  \advance\count0 by-1 \ifnum\count0>0\repeat}}
\def\tableau#1{\vcenter{\offinterlineskip
  \tableaustep=\tableauside\advance\tableaustep by-\tableaurule
  \kern\normallineskip\hbox
    {\kern\normallineskip\vbox
      {\gettableau#1 0 }%
     \kern\normallineskip\kern\tableaurule}%
  \kern\normallineskip\kern\tableaurule}}
\def\gettableau#1 {\ifnum#1=0\let\next=\null\else
  \squares{#1}\let\next=\gettableau\fi\next}

\newcommand{\comment}[1]{}
\newcommand{\Expect}[1]{\left\langle #1 \right\rangle}
\newcommand{\mbf}[1]{\mathbf #1}

\def\mpl{M_{\rm Pl}}
\newcommand{\be}{\begin{equation}}
\newcommand{\ee}{\end{equation}}
\newcommand{\bea}{\begin{eqnarray}}
\newcommand{\eea}{\end{eqnarray}}
\newcommand{\barr}{\begin{array}}
\newcommand{\earr}{\end{array}}

\def\Htwo{H^2{(t+\pi)}}
\def\Hdot{\dot H {(t+\pi)}}

\def\be{\begin{equation}}
\def\ee{\end{equation}}
\def\bea{\begin{eqnarray}}
\def\eea{\end{eqnarray}}

\def\mpl{M_{\rm Pl}}

\begin{document}

%\begin{titlepage}

\setcounter{page}{1} \baselineskip=15.5pt \thispagestyle{empty}

\begin{flushright}
%hep-th/yymmnnn\\
\end{flushright}
\vfil

\begin{center}

{\Large \bf (Small) Resonant non-Gaussianities:\\\vspace{0.1cm}  Signatures of a Discrete Shift Symmetry\\\vspace{0.4cm}  in the Effective Field Theory of Inflation}
\\[0.7cm]
{\large Siavosh R.~Behbahani~$^{a,b,c}$,  Anatoly Dymarsky~$^{d}$,\\\vspace{0.3cm}   Mehrdad Mirbabayi~$^{e}$ and Leonardo Senatore~$^{b,f}$}
\\[0.7cm]
%\vspace{.7cm}
%\vspace{.3cm}
{\normalsize { \sl $^{a}$ Theory Group, SLAC National Accelerator Laboratory, Menlo Park, CA 94025}}\\
\vspace{.2cm}

{\normalsize { \sl $^{b}$ Stanford Institute for Theoretical Physics, Stanford University, Stanford, CA 94306}}\\
\vspace{.2cm}

{\normalsize { \sl $^{c}$ Physics Department, Boston University, Boston, MA 02215}}\\
\vspace{.2cm}

{\normalsize { \sl $^{d}$  School of Natural Sciences, Institute for Advanced Study, Princeton, NJ 08540}}\\
\vspace{.2cm}

{\normalsize { \sl $^{e}$ CCPP, Physics Department, New York University, 4 Washington Place, New~York,~NY~10003}}\\
\vspace{.2cm}

{\normalsize { \sl $^{f}$ Kavli Institute for Particle Astrophysics and Cosmology, Stanford University and SLAC, Menlo Park, CA 94025}}\\
\vspace{.2cm}

\end{center}

\vspace{.8cm}

\hrule \vspace{0.3cm}
{\small  \noindent \textbf{Abstract} \\[0.3cm]
\noindent
We apply the Effective Field Theory of Inflation to study the case where the continuous shift symmetry of the Goldstone boson $\pi$ is softly broken to a discrete subgroup. This case includes and generalizes recently proposed String Theory inspired models of Inflation based on Axion Monodromy. The models we study have the property that the 2-point function oscillates as a function of the wavenumber, leading to oscillations in the CMB power spectrum. The non-linear realization of time diffeomorphisms induces some self-interactions for the Goldstone boson that lead to a peculiar non-Gaussianity whose shape oscillates as a function of the wavenumber. We find that in the regime of validity of the effective theory, the oscillatory signal contained in the $n-$point correlation functions, with $n>2$, is smaller than the one contained in the 2-point function, implying that the signature of oscillations, if ever detected, will be easier to find first in the 2-point function, and only then in the higher order correlation functions. Still the signal contained in higher-order correlation functions, that we study here in generality, could be detected at a subleading level, providing a very compelling consistency check for an approximate discrete shift symmetry being realized during inflation.}
 \vspace{0.3cm}
\hrule

\vfil
\begin{flushleft}
%\today
%March 20, 2008
\end{flushleft}

%\end{titlepage}

%\newpage
%\tableofcontents
\newpage

\section{Introduction}

In the light of the current experimental effort, it is important to explore all the possible signatures of Inflation. The Effective Field Theory (EFT) of Inflation~\cite{Cheung:2007st} is the ideal setup for doing this. In fact, as we will briefly review, by realizing that every inflationary model spontaneously breaks time-diff.s, it reduces the theory of the fluctuations for the most general model of  inflation to the one of the Goldstone boson associated with the breaking of time-diff.s.
As it often occurs with Goldstone bosons, the resulting Lagrangian is highly constrained by the symmetries and it allows for a complete phenomenological analysis. The EFT of Inflation has been developed in the following papers~\cite{Cheung:2007st,Cheung:2007sv,Senatore:2009gt,Senatore:2009cf,Senatore:2010jy,Creminelli:2006xe,Bartolo:2010bj,Bartolo:2010di,Senatore:2010wk,Creminelli:2010qf,senatore_complletion,Baumann:2011dt,Baumann:2011su,Baumann:2011nk,Nacir:2011kk}.
Many new observational signatures, such as the orthogonal shape of the 3-point function~\cite{Senatore:2009gt}, or the possibility of having large four-point function without a detectable 3-point function~\cite{Senatore:2010jy}, or the possibility to have Lorentz invariant (or conformal invariant) shapes of non-Gaussianities~\cite{Senatore:2010wk}, have been realized in this setup. The Effective Field Theory for Multifield Inflation has been realized in~\cite{Senatore:2010wk}, and dissipative effects induced by multiple degrees of freedom have been introduced in~\cite{Nacir:2011kk}. The use of Supersymmetry in the Effective Field Theory of Inflation has been introduced in~\cite{Senatore:2010wk,Baumann:2011nk}.

So far, the study of the EFT Lagrangian, both in the context of single clock and of multi-field, has been concentrated to the technically natural case where the interactions of the Goldstone are protected by an approximate continuos shift symmetry. The purpose of this paper is to generalize the study to the case where the continuos shift symmetry of the Goldstone boson is broken down to an approximate discrete shift-symmetry subgroup. This study is motivated by recent explicit stringy constructions of inflationary models called Axion Monodromy~\cite{McAllister}, where an axion plays the role of the inflaton and inflation occurs as the axion slowly rolls down its potential. This potential can be thought of as a standard slow-roll inflationary potential, such as for example $m^2\phi^2$, with  a small superimposed  oscillating component of the form $\cos(\phi/\Lambda)$. These models have the interesting observational consequence that
all the correlation functions oscillate as a function of the comoving wavenumber.

While the aformentioned setup relies on the explicit ultraviolet models being considered, here we will point out that such kind of self-interactions as well as generalizations thereof, can be thought of simply in terms of effective field theory and symmetries protecting the operators in the Lagrangian: an approximate continuous shift symmetry broken to an approximate discrete one. This has the advantage of disentangling what we might assume about the UV theory and what is actually connected to observations. It further allows us to explore all possible signatures in full generality. In the lucky event that we discover some non-Gaussianity or some oscillations in the cosmological data, we will therefore be able to connect observations to the essential Lagrangian of Inflation, which is the one of the Goldstone boson $\pi$. From there we might be able to move up in energy inferring about the UV completion of this Lagrangian.

In this paper we study the case where the Goldstone boson has an approximate discrete shift symmetry. In agreement with former literature, we find that the 2-point function has oscillatory features as we change the comoving wavenumber of the modes. Because of the non-linear realization of time-diff.s, these oscillatory features translate directly into interactions and therefore non-vanishing higher $n$-point functions. We systematically study this scenario, including limits where the 3-point function is very small and a relatively higher signal is in the 4-point function. However, contrary to former literature, we find that, within the validity of the effective theory, {\it i.e.} without the theory being strongly coupled ñor without the inclusion of new degrees of freedom, and avoiding tuning in the parameters, the signal in higher order $n$-point functions is {\it always} smaller than the one in the 2-point function. This implies that observational constraints or even detection for these kind of models must come from the 2-point function, with only marginal need to analyze higher order $n$-point functions, unless some (unexpected) degeneracies are present in the 2-point functions. The latter might indeed be detectable only at some subleading level, and such a detection would provide an extremely compelling consistency check of an approximate discrete shift symmetry for the Goldstone boson $\pi$.

\section{Effective Field Theory of Single-Clock Inflation and Approximate Shift Symmetry\label{sec:single-field}}

In this section we briefly review the effective action for single-clock inflation and we will describe how an approximate discrete shift symmetry for the clock field can be implemented within this approach. Readers familiar with the EFT of Inflation can skip directly to sec.~\ref{sec:discrte-symmetry}. The effective
action was developed in \cite{Cheung:2007st,Cheung:2007sv} and we refer the reader to those papers for a detailed explanation.
The construction of the effective theory is based on the following consideration. In a quasi de
Sitter background with only one dynamical degree of freedom, there is a privileged spatial slicing
given by the physical clock which allows us to smoothly connect to a decelerated hot Big Bang
evolution. The slicing is usually realized by a time evolving scalar $\phi(t)$, but this does not need necessarily to be the case. To describe perturbations
around this classical solution one can choose a gauge where the privileged slicing coincides with surfaces of
constant $t$, for example $\delta\phi(\vec x,t)=0$. In this `unitary' gauge there are no explicit scalar perturbations but only metric
fluctuations. As time diff.s have been fixed,
the graviton now describes one additional degree of freedom: the scalar perturbation has been eaten by the
metric. One therefore can build the most general effective action with operators that are functions
of the metric fluctuations and that are invariant under the linearly-realized time-dependent spatial
diff.s. As usual with effective field theories, this can be done in a low energy expansion
in fluctuations of the fields and derivatives. We obtain the following Lagrangian~\cite{Cheung:2007st,Cheung:2007sv}:
 \begin{eqnarray}
\label{eq:actiontad}\nonumber
S_{\rm E.H.\;+\; S.F.} & \!\!\!\!\!\!\!\!\!\!\!\!= \!\!\!\!\!\!\!\!\!& \!\!\!\int  \! d^4 x \; \sqrt{- g} \Big[ \frac12 M_{\rm
Pl}^2 R + M_{\rm Pl}^2 \dot H
g^{00} - M_{\rm Pl}^2 (3 H^2 + \dot H)+ \\\nonumber
&&+ \frac{1}{2!}M_2(t)^4(g^{00}+1)^2+\frac{1}{3!}M_3(t)^4 (g^{00}+1)^3+ \\
&& - \frac{\bar M_1(t)^3}{2} (g^{00}+1)\delta K^\mu {}_\mu
-\frac{\bar M_2(t)^2}{2} \delta K^\mu {}_\mu {}^2
-\frac{\bar M_3(t)^2}{2} \delta K^\mu {}_\nu \delta K^\nu {}_\mu + ... \Big]\ ,
\end{eqnarray}
where we denote by $\delta K_{\mu\nu}$ the variation of the extrinsic curvature of constant time surfaces with respect to the unperturbed FRW: $\delta K_{\mu\nu}=K_{\mu\nu}-a^2 H h_{\mu\nu}$ with $h_{\mu\nu}$ being the induced spatial metric, and where $M_{2,3}$ and $\bar M_{1,2,3}$ represent some time-dependent mass scales.

Let us comment briefly on (\ref{eq:actiontad}). The first term is the Eistein-Hilbert action. The first three terms are the only ones that start linearly in the metric fluctuations. The coefficients are such that when combined the linear terms in the fluctuations cancel. The action must start quadratic in the fluctuations. The terms in the second line start quadratic in the fluctuations and have no derivatives. The terms in third line represent higher derivative terms. Dots represent operators that start at higher order in the perturbations or in derivatives. This is the most general action for single clock inflation~\cite{Cheung:2007st}.

The unitary gauge Lagrangian describes three degrees of freedom: the two graviton helicities and
a scalar mode. This mode will become explicit after one performs a broken time diffeomorphism
(St\"uckelberg trick) to reintroduce the Goldstone boson which non-linearly realizes this symmetry. In analogy
with the equivalence theorem for the longitudinal components of a massive gauge boson \cite{Cornwall:1974km}, the Goldstone decouples from the two graviton helicities at high energies, and  the mixing can be neglected. As we will review and explicitly check later, it is possible to verify that in most situations of interest this is indeed the case and one can neglect the metric fluctuations.

As anticipated, we reintroduce the Goldstone boson ($\pi$) by performing a broken time-diff., calling the parameter of the transformation $-\pi$, and then declaring $\pi$ to be a field that under time diff.s of the form $t\rightarrow t+\xi^0(x)$ transforms  as
\be
\pi(x)\quad\rightarrow\quad \tilde\pi(\tilde x(x))=\pi(x)-\xi^0(x)\ .
\ee
In this way diff. invariance is restored at all orders. For example the terms containing $g^{00}$ in the Lagrangian  give rise to the following terms:
\be
g^{00}\quad\rightarrow \quad\frac{\partial (t+\pi)}{\partial x^\mu}\frac{\partial (t+\pi)}{\partial x^\nu}g^{\mu\nu} \quad\rightarrow\quad g^{00} +2 g^{0\mu} \partial_\mu \pi + (\partial \pi)^2 .
\ee
We refer to~\cite{Cheung:2007st} for details.
If we are interested just in effects that are not dominated by the mixing with gravity, then we can neglect the metric perturbations and just keep the $\pi$ fluctuations. In this regime, a term of the form $g^{00}$ in the unitary gauge Lagrangian becomes:
\be
g^{00}\quad\rightarrow\quad -1-2\dot\pi-\dot\pi^2+\frac{1}{a^2}(\partial_i\pi)^2\ .
\ee
Furthermore, we can assume that the $\pi$ has an approximate continuous shift symmetry, which becomes exact in the limit when the space time is exactly de Sitter \cite{Cheung:2007st}. This allows us to neglect terms in $\pi$ without a derivative that are generated by the time dependence of the coefficients in (\ref{eq:actiontad}).
Implementing the above procedure in the Lagrangian of (\ref{eq:actiontad}), we obtain the rather simple result:
\begin{eqnarray}\label{eq:Spi}
S_{\rm \pi} =\int d^4 x   \sqrt{- g} \left[ -M^2_{\rm Pl}\dot{H} \left(\dot\pi^2-\frac{ (\partial_i \pi)^2}{a^2}\right)
%- M^2_{\rm Pl} \left(3H^2 +\dot{H}\right)+ \right.\\
+2 M^4_2
\left(\dot\pi^2+\dot{\pi}^3-\dot\pi\frac{(\partial_i\pi)^2}{a^2}
\right) -\frac{4}{3} M^4_3 \dot{\pi}^3 +\ldots\right] \ , \nonumber \\
\eea
where for simplicity we have neglected the terms originating from the extrinsic curvature as they are usually only important for inflationary models where the space time is very close to de-Sitter space \cite{Cheung:2007st}.

We notice that when $M_2$ is different from zero the speed of sound of the fluctuations is different from one. We have the following relation:
\be\label{eq:M2cs}
M_2^4=-\frac{1-c_s^2}{c_s^2}\frac{\mpl^2\dot H}{2}\ .
\ee
At leading order in derivatives, there are two independent cubic self-interactions, $\dot\pi(\partial_i\pi)^2$ and $\dot\pi^3$, which can induce detectable non-Gaussianities in the primordial density perturbations. A small speed of sound (i.e.~a large $M_2$) forces large self-interactions of the form $\dot\pi(\partial_i\pi)^2$, while the coefficient of the operator $\dot\pi^3$ is not fixed because it also depends on $M_3$. Cosmological data can therefore constrain (or measure) the parameters of the above Lagrangian. This approach has been recently applied to the WMAP data in \cite{Senatore:2009gt}, giving constraints on $M_2$ and $M_3$, as well as on the higher derivative operators that we have omitted in (\ref{eq:Spi}). This is exactly analogous to what happens for data from particle accelerators when the precision electroweak tests of the Standard Model are carried out \cite{Peskin:1991sw,Barbieri:2004qk}.

\subsection{Approximate discrete shift symmetry\label{sec:discrte-symmetry}}

When above we obtained eq.~(\ref{eq:Spi}), we explicitly assumed that there was an approximate continuous shift symmetry for $\pi$. This allowed us to neglect all the terms that would result from the time dependence of the coefficients present in the action, as, when we reinsert $\pi$ from unitary gauge, any function of time becomes function of $\pi$, explicitly:
\be
f(t)\rightarrow\ f(t-\pi)\simeq f(t)-\dot f \pi+\ldots\ .
\ee
We notice that neglecting these terms has nothing to do, at least in principle, with taking the decoupling limit. It is simply a technically natural assumption of imposing this symmetry on $\pi$. However, it is conceivable that during inflation this symmetry could be broken. This is so because inflation needs to be a phase of quasi de-Sitter space once averaged on time scales of order $H$. This is indeed the necessary requirement for solving the horizon problem, and also to produce quasi scale-invariant perturbations, at least in each reasonable bin of wavenumbers. Scale invariant perturbations are a consequence of time translation invariance during inflation, and it is conceivable that time translations could be broken on short time scales and be approximately recovered on long time scales, to give rise to experimentally acceptable quasi scale-invariant perturbations. On smaller time scales, the space time can be quite different from de Sitter, and still give an acceptable model of inflation that predicts scale invariant perturbations.

Since time translation invariance of the background is mapped into a shift symmetry of $\pi$, this discussion leads us to explore the possibility of relaxing the continous shift symmetry for $\pi$. It is in general very hard to protect the lightness of a scalar field, and shift symmetries represents a natural way to do this~\footnote{Another possibility is Supersymmetry, but in general it is not powerful enough in an expanding universe.}. Breaking the shift symmetry in general means that there are strong corrections to the mass of the fields that make its lightness fine tuned. However, there is at least one known exception to this. This is the case when the field has a violently broken continuous shift symmetry, which however leaves a softly broken discrete shift symmetry.
Let us consider for example a scalar field in Minkowski space with a Lagrangian
\be
S=\int d^4x\, \left[(\partial\phi)^2+\mu^4\cos(\phi/F)\right] \ .
\ee
This Lagrangian is typical for axions.  This theory is non-renormalizable with unitarity bound equal to
\be\label{eq:unitarity_bound}
\Lambda_U\simeq 4\pi F\ .
\ee
In order for the particle to have mass smaller than the unitary bound, we need to impose $\mu\lesssim F$. The parameter $\mu$ softly breaks the continuous shift symmetry of $\phi$ leaving out an unbroken discrete shift symmetry
\be
\phi\quad\rightarrow\quad \phi+2\pi F \ .
\ee
Notice that any interaction breaking the continuos shift symmetry is proportional to $\mu^4$. Possibly there could be operators of the form $\sim(\partial\phi)^n/F^{2n-4}$ or other operators induced by loops of the potential terms that are compatible with the continuos shift symmetry and therefore cannot renormalize operators that break it. This means that $\mu\ll F$ is technically natural. The symmetry pattern in this case is such that at energies $\mu\ll E\ll F$, the continuous shift symmetry is approximate, while the discrete shift symmetry is always exact down to energies comparable to the mass of the particle $\mu^2/F$, which is the lowest energy scale at which the effective theory makes sense. 

Let us now softly break the discrete shift symmetry. Soft breaking means that the radiative corrections induced by the terms breaking the symmetry are small and leave the original symmetry to be a good approximate one. Let us call the scale suppressing the operators breaking the discrete shift symmetry $F_d$. Examples of such terms could be a term as $\phi^6/F_d^2$. In this case we need to ensure that the loop-induced potential operators from this term are smaller than the mass induced by the oscillatory potential. This implies the constraint $F_d\gtrsim F^3/\mu^2$. Another possibility for softly breaking the discrete shift symmetry is by adding a potential of the form $\mu^{4}_d\cos(\phi/F_d)$, which is stable under radiative corrections. Here the condition that the original discrete shift symmetry is a good approximate one translates into $F_d\gtrsim F$ for $\mu_d\lesssim \mu$ or $F_d\gtrsim F \mu_d^2/\mu^2$ for $\mu_d\gtrsim \mu$. This is a pattern of symmetry breaking when a continuos shift symmetry is softly broken to a discrete one that is in turn is softly broken.

The purpose of this paper is to apply the above pattern of symmetry breaking to the Effective Field Theory of Inflation. It is technically natural for the Goldstone boson $\pi$ to have an approximate continuous shift symmetry. However, as we just discussed, it is also technically natural to have a softly broken discrete shift symmetry.  The simplest scenario in which this can be realized is obtained by setting to zero the coefficient of all higher order operators in unitary gauge (we will include them later) and taking $H(t)$ and $\dot H(t)$ to have small oscillating components:
\beq
\label{effaction}
S=\int d^4x \sqrt{-g}\mpl^2\;\left[\frac{1}{2}R
         -\left(3H^2(t+\pi)+\dot H(t+\pi)\right) +\dot H(t+\pi)\left(-(1+\dot\pi)^2  +(\partial_i\pi)^2\right)\right]\, . \nonumber \\
\eeq
Here we have neglected metric fluctuations. We will come back later to this point arguing that they are irrelevant.
The original models considered in the literature \cite{Chen1,Chen2,Bean,Silverstein,McAllister,Flauger1,Flauger2,Pajer} where a small sinusoidal term has been added to the slow-roll potential
\beq
\label{potential}
S=\int d^4x\,\sqrt{-g}\; \left[\frac{1}{2} (\partial\phi)^2-V_{\rm sr}(\phi)-\mu^4\cos(\phi/F)\right]\,,
\eeq
are included into this category. Here $V_{\rm sr}$ is the slow roll potential breaking the discrete shift symmetry of the inflaton $\phi$. 

The mentioned pattern of symmetry breaking is realized if we imagine that the Hubble scale is the superposition of a slowly time-dependent function and a small rapidly oscillating function:
\be\label{eq:Hsocillatin}
H(t)=H_{\rm sr}(t)+H_{\rm osc}(t) \sin(\omega t) .
\ee
Here $H_{\rm sr}(t)$ and $H_{\rm osc}(t)$ are slowly time dependent. Their time dependence is parametrized by the slow roll parameter $\epsilon$
\be
\epsilon=-\frac{\dot H}{H^2}\sim \frac{\dot H_{\rm sl}}{H^2_{\rm sl}}\sim \frac{\dot H_{\rm osc}}{H^2_{\rm osc}}\ll 1\ ,
\ee
which is the parameter controlling the smallness of the breaking of the discrete shift symmetry. 
In order to have inflation, we require $H_{\rm osc}\ll H_{\rm sr}$. Indeed $\epsilon$ controls the breaking of the discrete shift symmetry. The continuous shift symmetry is broken by $H_{\rm osc}$ to a discrete one $\pi(\vec x,t)\rightarrow \pi(\vec x,t)+2\pi/\omega$. We are interested in the regime where the discrete shift symmetry is softly broken, which means that the unitarity bound induced by the operators that are compatible with the discrete symmetry is smaller than the one induced by the operators that break it. If we substitute back $H(t+\pi)$ into the action, and Taylor expand to obtain the leading operators, we will easily see that the relative weight of operators compatible with the discrete symmetry to the ones not respecting it is the ratio of the time derivative of the oscillating and non oscillating parts of $H$. Requiring therefore that the cutoff from the shift-symmetry-respecting operators $\sim F\simeq (-2\dot H\mpl^2)^{1/2}/\omega$ is smaller than the unitarity bounds induced by the gravity induced interactions ($\sim\mpl$) and by the slow-roll mediated interactions, we obtain the condition
\be
\label{alpha}
\alpha\equiv\frac{\omega}{H}\gg \epsilon^{1/2}\ .
\ee

It is worth highlighting the hierarchy among the various components of $H$ as the various powers of the time derivatives are considered. As we mentioned before, the non-oscillating part of $H$ is dominant. This must be the case also for $\dot H$ as otherwise the sign of $\dot H$ would change and the Goldstone boson $\pi$ would become a ghost, leading to a catastrophic instability. This implies $\epsilon\gtrsim \alpha\; H_{\rm osc}/H$ and suggests us to define
\be\label{eosc}
\epsosc\equiv\frac{H_{\rm osc}\alpha}{H\epsilon}\ ,\qquad |\epsosc|<1\ .
\ee
 The oscillatory term can potentially dominate only starting at the level of $\ddot H$. 
 
%%%%%%%%%%%%%%%%%%%%%%%%%%%%%%%%%%%%%%%
\subsection{Two-point function}

Let us now compute the 2-point function. In the decoupling limit where we neglect metric fluctuations, the Taylor expansion of the action in \eqref{effaction} reads
\bea
\label{act_n}
S_n=&&\int d^4x\, a^3\, \mpl^2 \left[-\frac{3}{n!}\partial_t^n(H^2)\pi^n
     -\frac{2}{n!}H^{(n+1)}\pi^n
     -\frac{2}{(n-1)!}H^{(n)}\pi^{n-1}\dot\pi \right.\nonumber\\
&&\qquad\qquad\qquad\left.  -\frac{1}{(n-2)!}H^{(n-1)}\pi^{n-2}(\dot\pi^2-(\partial_i \pi)^2)\right]=\\ \nonumber
=&&\int d^4x\, a^3\, \mpl^2 \left[
 -\frac{1}{(n-2)!}H^{(n-1)}\pi^{n-2}(\dot\pi^2-(\partial_i\pi)^2)+\frac{3}{n!}(2HH^{(n)}-\partial_t^n(H^2))\pi^n\right]\ ,
\eea
where in the third line we have integrated by parts $\pi^{n-1}\dot\pi$ and $H^{(n)}=\partial^n_t H$. We will justify later having taken the decoupling limit.
The linear equation of motion reads
\beq
\label{lineq}
\ddot\pi+(3H+\ddot H/\dot H)\dot\pi - \partial^2 \pi=0\,,
\eeq
where we neglected a term in $\epsilon=-\dot H/H^2$. This also will be justified shortly.
Introducing the conformal time $\tau\equiv\int dt/a$, and defining $x\equiv -k\tau$, eq.~(\ref{lineq}) can be written as
\beq
\label{lineqx}
\partial_x^2\pi_{\bf k}-\frac{2(1+\delta)}{x}\partial_x\pi_{\bf k}+\pi_{\bf k}=0\,,
\eeq
where $\delta\equiv\ddot H/(2H\dot H)$. This equation is very similar to the one found in \cite{Flauger1,Flauger2} for the variable $\delta\phi$ and can be solved in a very similar way. So we delay the derivation to Sec.~\ref{mode_function}. The important point is that $\delta$ oscillates with time. As modes start deep inside the horizon and redshift up to Hubble, their proper frequency crosses $\omega$ before Hubble crossing if $\alpha\gtrsim1$. In this case the modes go through a resonance and the state becomes different from the Bunch-Davies vacuum while still well inside the horizon. In the $\alpha\gg1$ case the solution to the linear equation can be found in the saddle point approximation and by expanding at linear order in $\epsosc$. We will shortly see that $\epsosc\ll1$ for observational constraint. The effect on the power spectrum of $\R=-H\pi+O(\pi^2)$ is given by
 \bea
  \label{2pt}
\Expect{\R_\mbf{k}\R_\mbf{{k'}}} & = & (2\pi)^3 \delta^{(3)}(\mbf{k}+\mbf{k'}) \frac{H^2(t_\star)}{4\ep(t_\star) \mpl^2 k^3} \left[ 1- \left(\frac{\pi}{2}\right)^{1/2}\epsosc\, \alpha^{1/2}\sin\left(\alpha\ln(2k/k_\star)\right)\right]\,, \qquad \alpha\gg 1, \nonumber \\
  \eea
  where $t_\star$ is such that $k/a(t_\star)=H(t_\star)$ and we have assumed that the time of the resonance and the time of horizon crossing are not too far, so that variations of $H_{\rm sr}\, , \ldots$ between the two times are negligible. Here and in the rest of the paper $k_\star$ is a comoving wavenumber representing the physical phase of the oscillating term.  The leading effects from the slow roll parametes have been included with the standard treatment of evaluating all quantities at horizon crossing.  This result agrees with the one in~\cite{Flauger1,Flauger2}~\footnote{In order to compare, notice that $\epsosc=-6\, b_{there}/\alpha$. We prefer to use  $\epsosc$ rather than $b$ because, as we stressed in the introduction, we find this notation to be more justified in terms of symmetries. Indeed, the constraint $b<1$ that is often assumed in the literature comes from imposing the monotonicity of the potential for standard slow roll inflationary models. Such a constraint is  not necessary.}.

Notice that the effect of oscillations appears as a modification of the tilt: a sort of oscillating tilt. Constraints on the tilt imply that (see~\cite{Flauger1} for a detailed analysis)
\be\label{eq:epsoscv}
\epsosc\alpha^{1/2}\lesssim 10^{-2}\sim \epsilon \qquad\Rightarrow\qquad \epsosc\lesssim\epsilon\sim 10^{-2}\ll1  \ , \quad {\rm for} \quad \alpha\gg1\ ,
\ee
which justifies our assumption of expanding at linear level in $\epsosc$. In the case $\alpha\lesssim 1$, the requirement of soft breaking of the discrete shift symmetry implies $\alpha\gtrsim \epsilon^{1/2}$, which combined with (\ref{eq:epsoscv}) leads to the bound
\be
\epsosc\lesssim \epsilon^{3/4}\ll 1\ .
\ee
We can therefore treat $\epsosc$ as an expansion parameter much smaller than one for all values of $\alpha$.

In the case $\alpha\ll 1$, the oscillations of the background is slow with respect to Hubble, and the modes do not undergo through a resonance before freezing out. In this case the effect of the oscillations shows up in the value of $H$ and $\epsilon$ at horizon crossing, leading to a small and slowly oscillating tilt. The result reads (see also~\cite{Kobayashi:2010pz}):
\beq
  \label{2pt2}
\Expect{\R_\mbf{k}\R_\mbf{{k'}}} & = & (2\pi)^3 \delta^{(3)}(\mbf{k}+\mbf{k'}) \frac{H^2(t_\star)}{4\mpl^2\ep(t_\star) k^3}\,, \qquad  \alpha\ll 1 \ .
  \eeq

%%%%%%%%%%%%%%%%%%%%%%%%%%%%%%%%%%%%%%%

\subsection{$n$-point correlator}\label{simplest}

In order to evaluate the single vertex connected $n$-point correlation function we need to derive the $n$-th\, order interaction Lagrangian for $\pi$.
Since derivatives of the background carry potentially large factors of $\alpha$ (we will indeed see that the most interesting physics happens when $\alpha\gg1$), it is useful to make derivatives of the background quantities as explicit as possible by integrating by parts the action (\ref{effaction}). We obtain
\beq
\label{Sn2}
S_n&=&\int d^4x a^3 \mpl^2
    \left[\frac{1}{(n-1)!}H^{(n)}\pi^{n-1}\dot\pi
        +\frac{3}{(n-1)!}HH^{(n-1)}\pi^{n-1}\dot\pi \right.\nonumber\\
&&\qquad \left. +\frac{1}{(n-1)!}H^{(n-1)}\pi^{n-1}\ddot\pi
        -\frac{1}{(n-1)!}H^{(n-1)}\pi^{n-1}\Delta\pi+\frac{3}{n!}(2HH^{(n)}-\partial_t^n(H^2))\pi^n\right]\nonumber\\
\!\!\!\! \!\!\!\! \!\!\!\! \!\!\!\! \!\!\!\! \!\!\!\! \!\!\!\! \!\!\!\! \!\!\!\! \!\!\!\! \!\!\!\! \!\!\!\! \!\!\!\! \!\!\!\! \!\!\!\!  &\simeq&\int d^4x \left[
 -  \frac{a^3 \mpl^2}{n!}\left(H^{(n+1)}+3 HH^{(n)} \right)\pi^n +f_n(\pi)   \left.\frac{\delta{\cal L}}{\delta \pi}\right|_1\right]+{\cal O}(\epsosc^2) \ , 
\eeq
where
\beq
f_n(\pi)\equiv \frac{1}{2(n-1)!}\frac{H^{(n-1)}}{\dot H}\pi^{n-1}\,,
\eeq
and  $ \left.\delta{\cal L}/\delta \pi\right|_1$ is the linearized equation of motion. Here we have taken the leading order contribution in $\epsosc$, and used the following scaling for the time derivatives of $H$ from (\ref{eq:Hsocillatin}):
\bea\label{eq:Hosc}
&& \dot H\sim \epsilon\, H^2\\ \nonumber
&& \partial_t^n H\sim \epsilon\,\epsosc \alpha^{n-1}H^{n+1} \sin(\omega t) \quad {\rm for}\quad n\geq 2\ .
\eea
 The second term in the third line of (\ref{Sn2}) is suppressed with respect to the first by a factor $1/\alpha$. The terms in the last line proportional to the linear equation of motion of can be removed through the following field redefinition
\beq
\label{redefine}
\pi = \pi_r - \sum_n f_n(\pi_r)\,.
\eeq
This has the side effect of modifying the interaction Lagrangian ${\cal L}_n$ for all $n\geq 4$. However, the terms induced by the field redefinition are suppressed by at least one power of $\epsosc\ll 1$ with respect to the leading interaction and so are negligible at the leading order. Nonetheless in conversion of the correlators of $\pi_r$ to those of $\pi$ these field redefinitions must be in principle taken into account as we will do later in this section. We are therefore left with the following leading interaction:
\beq
\label{Sn3}
S_n =-\int d^4x\, a^3\;
       \frac{1}{n!}\mpl^2 \left(H^{(n+1)}+3 H H^{(n)}\right)\pi_r^{n}+{\cal O}(\epsosc^2)\,.
\eeq
After canonical normalization of $\pi_r$
\beq
\label{pi_c}
\pi_c\equiv (-2\mpl^2\dot H)^{1/2}\pi_r\,,
\eeq
the interaction Lagrangian becomes
\beq
\label{Ln_pi}
{\cal L}_n(\pi_c)=-\frac{1}{n!} \mpl^2 \left(H^{(n+1)}+3HH^{(n)}\right)
               \left(\frac{\pi_c}{(-2\mpl^2\dot H)^{1/2}}\right)^n\,.
\eeq
It is possible to show that this Lagrangian agrees with the $n^{\rm th}$ order interaction Lagrangian obtained in \cite{Pajer} where a model of a slowly rolling scalar field was considered~\footnote{We will show the equivalence just at leading order in $\epsosc$. The interaction Lagrangian of order $n$ in \cite{Pajer}  reads
\beq
\label{Ln_phi}
{\cal L}_n(\delta\phi) =-\frac{1}{n!}V^{(n)}(\phi)(\delta\phi)^n\,.
\eeq
Using the Friedmann equations that allows us to write $V(\phi)$ and $\dot\phi$ in terms of $H$ and its derivatives
\beq
V=\mpl^2(\dot H +3 H^2)\,,\qquad \quad
\label{phi_dot}
\dot\phi^2&=&-2 \mpl^2\dot H\,,
\eeq
the derivatives of $V$ with respect to $\phi$ can be written in terms of time derivatives of $H$
\beq
V'=\mpl^2\frac{1}{\dot\phi}\partial_t(\dot H +3 H^2)
   = \mpl^2\frac{1}{\dot\phi}(\ddot H +6 H\dot H)\,,\\
V''= \mpl^2\frac{1}{\dot\phi}
     \partial_t\left(\frac{\ddot H +6 H\dot H}{\dot\phi}\right)
   =  \mpl^2\frac{1}{\dot\phi^2}(\dddot H +3 H\ddot H)\,.
\eeq
It is easy to see that in higher derivatives than $V''$, taking derivatives of $\dot\phi$ results in second or higher powers of $\epsosc$. We obtain
\beq
\label{V_n}
V^{(n)}=\mpl^2\frac{1}{\dot\phi^n}\left(H^{(n+1)} +3 H H^{(n)}\right)+{\cal O}(\epsosc^2)\,,
\eeq
and finally using \eqref{phi_dot} and \eqref{Ln_phi}
\beq
{\cal L}_n(\delta\phi) =-\frac{1}{n!}\mpl^2\left(H^{(n+1)} +3 H H^{(n)}\right)
         \left(\frac{\delta\phi}{(-2\dot H \mpl^2)^{1/2}}\right)^n\,,
\eeq
which is identical to the $n^{\rm th}$ order Lagrangian \eqref{Ln_pi} for canonically normalized $\pi_c$ at leading order in $\epsosc$.}. The calculation is straightforward following~\cite{Pajer}. The Fourier modes undergo resonance at $k/a(t)\sim \omega$ and the integral can be done in the saddle point approximation. Using the conversion $\R=-H\pi$, we obtain
\beq
\label{R_N}
\Expect{\prod_{i=1}^n\R_{\mathbf{k}_i}}&=&  (2\pi)^3\delta^3\left(\sum_{i=1}^n \mathbf{k}_i\right) A_nB_n(k_i)\,,\nonumber\\
 A_n &\equiv& (-)^{n+1} \frac{ \epsosc \sqrt{2\pi}}{4} \alpha ^{2n-7/2}\left(\frac{H^2}{4\ep\mpl^2}\right)^{n-1}\,,\\
B_n(k_i)&\equiv& \frac{1}{ K^{n-3} \prod_i k_i^2} \left[ \sin\left(\alpha \ln(K/k_\star)\right)+\frac1\alpha \cos\left(\alpha \ln (K/k_\star)\right) \sum_{j,i} \frac{k_i}{k_j} \right]\,,\nonumber
\eeq
where $K\equiv\sum_ik_i$. These expressions are valid only for $\alpha\gg 1$ and they agree with \cite{Pajer}. 

The result for $\alpha\sim 1$ or smaller can be estimated by taking $\alpha\rightarrow 1$ in (\ref{R_N}). In this case we get an effective $f_{\rm NL}\sim \epsosc$, which, after including the bound from the tilt of the 2-point function, results in a negligible amount of non-Gaussianity. We therefore safely restrict to the case $\alpha\gtrsim 1$ for the rest of the paper.  We will explain in App.~\ref{app:mixing}~and~\ref{app:redefinition} the reason why coupling to gravity can be neglected and the linear conversion between $\zeta$ and $\pi$ is sufficient.

Let us first comment on the scaling of \eqref{R_N} with $\alpha$. It is clear from \eqref{Ln_pi} that as one goes to higher order interactions, at each level there is an extra time derivative which gives a factor of $\omega=\alpha H$. Moreover since $n^{\rm th}$ order interaction Hamiltonian contains $n$ fields with a wavefunction of the form
\beq
\label{BD}
f_k(\tau)\propto \frac{H}{\sqrt{2k^3}}(1+ik\tau)e^{-ik\tau}\,,
\eeq
in view of the fact that the resonance happens around $k/a\sim \omega$ (or equivalently $|k\tau|\sim \alpha$), we expect to get two more factors of $\alpha$ at each order of $\Expect{\R^n}$.

It is easy to get the $n$ dependence of the formula above. We just need to realize that the ratio  of $\langle\zeta^n\rangle/\langle\zeta^2\rangle^{n/2}$ scales as the ratio of the interacting Lagrangian ${\cal L}_n$  and the quadratic Lagrangian ${\cal L}_2$ evaluated at energy scale of order $\omega$ where the resonance happens:
\beq\label{eq:scaling}
\frac{\langle\zeta^n\rangle}{\langle\zeta^2\rangle^{n/2}}\sim \left.\frac{{\cal L}_n}{{\cal L}_2}\right|_{E\sim \omega} \ .
\eeq
Notice the important difference here that the non-Gaussianities are dominated by interactions happening when the energy is of order $\omega$ contrary to the standard case when non-Gaussianities are dominated by interactions happening at energy of order $H$. This difference arises due to the resonance that happens when the  mode is way inside the horizon. Plugging into (\ref{eq:scaling}) and using~(\ref{eq:Hosc}) we obtain
\bea\label{eq:scaling2}
\frac{\langle\zeta^n\rangle}{\langle\zeta^2\rangle^{n/2}}\sim\epsosc \left(\omega \pi\right)^{n-2}\sim\epsosc \left(\alpha^2\zeta\right)^{n-2}\ ,
\eea
where we have used that $\pi$ at an energy scale $E$, $\pi_E$, is related to $\pi$ at Hubble by $\pi_E\sim (E/H)\pi_H\sim E/H^2\zeta$ and that $\zeta\sim  H\pi_H$. To get (\ref{R_N}), there is another factor of $\alpha^{1/2}$ that comes from the saddle point approximation. More instructively, (\ref{eq:scaling}) can be also re-written as
\bea\label{eq:scaling3} \nonumber
\frac{\langle\zeta^n\rangle}{\langle\zeta^2\rangle^{n/2}}&\sim& \epsosc\,\alpha^{1/2} \left(\omega \pi\right)^{n-2}\sim\epsosc\,\alpha^{1/2} \left(\frac{\omega}{(-\dot H\mpl^2)^{1/2}}\pi_c\right)^{n-2}\sim\epsosc\,\alpha^{1/2} \left(\frac{\pi_c}{F}\right)^{n-2}\\ 
&\sim&\epsosc\,\alpha^{1/2} \left(\frac{\omega}{F}\right)^{n-2}\ .
\eea
where $F$ is related to the unitarity bound $\Lambda_U$ of the theory due to the discrete-symmetry preserving terms by the relationship $\Lambda_U\simeq 4\pi F$. 
Performing the same procedure keeping track of numerical factors, we find the value of $F$ to be
\be
\label{f}
\cos\left(\omega (t+\pi)\right)=\cos\left(t+\frac{\omega}{(-2\dot H\mpl^2)^{1/2}}\pi_c\right)\qquad\Rightarrow\qquad F\simeq \frac{\left(-2\dot H\mpl^2\right)^{1/2}}{\omega}\ .
\ee

The perturbative series is defined as long as $\langle\zeta^n\rangle/\langle\zeta^2\rangle^{n/2}\ll1$, which implies $\omega\ll \Lambda_U$. This obviously corresponds to the case in which the resonance frequency is smaller than the unitarity bound of the theory: $\omega\ll\Lambda_U$~\footnote{Alternatively we could say that in the case of $\omega\gg F$, we could still perform the calculation of the $n$-point function but we should restrict ourself to frequencies much smaller than $F$. In this case the effect of the oscillatory functions would be suppressed by $F/\omega$ instead of leading to an enhancement.}. Thus the bound on the consistency of the effective theory implies that
\be\label{eq:alphaboundprec}
\alpha^2 \ll \frac{4\pi}{\sqrt2\Rtwo^{1/2}}\ .
\ee
This is a very important bound that will have relevant observational consequences. Indeed eq.~(\ref{eq:scaling2}) is proportional to the signal to noise, and we see that within the validity of the effective theory the lowest $n$-order correlation functions have the leading signal to noise. Before being sure that this is the case, we need to take care of the mixing with gravity and of the redefinition between $\zeta$ and $\pi$, and be sure that terms that we neglected do not represent an important contribution for higher order $n$-point functions. We prove that these contributions are negligible respectively in App.~\ref{app:mixing}~and~\ref{app:redefinition}. This allows us to conclude that as a general prediction of the effective field theory the leading signal to noise is in the 2-point function. It would be interesting to investigate if this conclusion can change if one considers specific UV completions of the Effective Theory where new degrees of freedom are included so that resonance frequencies larger than $F$ can be consistently considered.

While a complete answer can only come from detailed studies of specific UV completions, we can anticipate that such a possibility seems unlikely to us. It seems indeed very hard to induce resonance effects at frequencies $\omega\gtrsim F$, even after a UV completion has soften the amplitudes to preserve Unitarity at such energies, as in this case it is expectable that the oscillating  components of the background should be  erased by the vacuum quantum fluctuations of the fields when the modes have frequencies of order~$\omega$. On top of this, given that the role of the UV completion is to soften the amplitudes, it seems even harder the possibility that the eventual remaining signal would be stronger in the higher order $n$-point rather than in the 2-point function.

%%%%%%%%%%%%%%%%%%%%%%%%%%%%%%%%%%%%%%%%%%%%%%%%%%%%

%%%%%%%%%%%%%%%%%%%%%%%%%%%%%%%%%%%%%%%%%%%%%%%%%%%%
%%%%%%%%%%%%%%%%%%%%%%%%%%%%%%%%%%%%%%%%%%%%%%%%%%%%
%%%%%%%%%%%%%%%%%%%%%%%%%%%%%%%%%%%%%%%%%%%%%%%%%%%%

%%%%%%%%%%%%%%%%%%%%%%%%%%%%%%%%%%%%%%%%%%%%%%%%%%%%
%%%%%%%%%%%%%%%%%%%%%%%%%%%%%%%%%%%%%%%%%%%%%%%%%%%%
%%%%%%%%%%%%%%%%%%%%%%%%%%%%%%%%%%%%%%%%%%%%%%%%%%%%

%%%%%%%%%%%%%%%%%%%%%%%%%%%%%%%%%%%%%%%%%%%%%%%%%%%%%%%%%%%%%%%%
%%%%%%%%%%%%%%%%%%%%%%%%%%%%%%%%%%%%%%%%%%%%%%%%%%%%%%%%%%%%%%%%

\section{Upper Bound and Observability}
%%%%%%%%%%%%%%%%%%%%%%%%%%%%%%%%%%%%%%%%%%%%%%%%%%%%

The upper bound (\ref{eq:alphaboundprec}) from the consistency of the effective theory restricts the resonant $n$-point function amplitudes  $A_N$ in \eqref{R_N} to satisfy
\beq
\label{hierarchy}
\frac{|A_2|}{\Rtwo}\gsim\frac{|A_3|}{\Rtwo^{3/2}}\gsim\cdots\gsim\frac{|A_n|}{\Rtwo^{n/2}}\gsim\cdots\ .
\eeq
This implies that the resonant part of the 2-point function is the best observable prediction of this model, unless some (unexpected) degeneracies in the 2-point function are present~\footnote{In~\cite{Flauger1}, an analysis of the 2-point function was carried out for the CMB WMAP data, identifying a degeneracy with $\Omega_b$, leading to weaker limits of at most a factor of 5 for some small region of values of $\alpha$ smaller than $10$, when a specific value of the phase is chosen. These are much smaller values for $\alpha$ (by a factor of at least 10) than the maximum allowed ones. Since the signal in the 3-point function scales relative to one in the 2-point function as the square of the ratio of the $\alpha$'s, we conclude that this small degeneracy cannot be broken by analysis of higher order $n$-point functions. Further this degeneracy will most probably be broken completely by the Planck data or already at present by carrying out an analysis using the data from ACT and SPT where a higher number of acoustic peaks are measured with a high signal to noise.}. This is one of the most important results of this paper.

To derive this statement more rigourously,  let us estimate the precision of a measurement of an $n$-point function using the Fisher matrix method (see for instance \cite{Komatsu}). Suppose we measure $N_{pix}$ points in a volume $V$ of the sky. Transforming to momentum space, it amounts to observing modes up to a $k_{max}$ momentum, where $Vk_{max}^3\sim N_{pix}$. The ratio of signal to noise can then be estimated as follows
\beq
\label{S/N}
\left(\frac{S}{N}\left(\Expect{\R^n}\right)\right)^2&\sim& V^n\int \frac{d^3\mbf{k}_1}{(2\pi)^3}\cdots \frac{d^3\mbf{k}_n}{(2\pi)^3}\frac{\Expect{\R_{\mbf{k}_1}\cdots\R_{\mbf{k}_n}}\Expect{\R_{\mbf{k}_1}\cdots\R_{\mbf{k}_n}}}{\Expect{\R_{\mbf{k}_1}\cdots\R_{\mbf{k}_n}\R_{\mbf{k}_1}\cdots\R_{\mbf{k}_n}}} 
\eeq
where for simplicity we have neglected the transfer functions and we have taken a 3-dimensional survey. Conclusions are not expected to depend qualitatively on these simplifications. Taking $(2\pi)^3\delta^3(\mbf{0})=V$ and neglecting for the moment  logarithmic corrections, we obtain
\beq
\left(\frac{S}{N}\left(\Expect{\R^n}\right)\right)^2\sim \frac{A_n^2}{\Rtwo^{n}} V \int \frac{d^3\mbf{k}_1}{(2\pi)^3}\cdots \frac{d^3\mbf{k}_{n-1}}{(2\pi)^3}\frac{1}{K^{2n-6}\prod_ik_i}\sim \frac{A_n^2}{\Rtwo^{n}} N_{pix}\,,
\eeq
which after combining with \eqref{hierarchy} results in
\beq
\frac{S}{N}\left(\delta\Expect{\R^2}\right)>\frac{S}{N}\left(\Expect{\R^3}\right)>\cdots >\frac{S}{N}\left(\Expect{\R^N}\right)>\cdots\,.
\eeq
A careful computation, taking into account all numerical factors, confirms that the ratio of the signal to noise for the 3-point function compared to the one from the 2-point function reaches the highest value of $0.87<1$ even when we push $\alpha$ to be so high to saturate the bound of~\eqref{eq:alphaboundprec} (see Appendix \ref{signal}). For consistency of the perturbative expansion, $\alpha$ should be much smaller than $\alpha_{\rm saturation}$, and the ratio of the signal from the 3-point function with respect to the one from the 2-point function gets suppressed by the ratio $(\alpha/\alpha_{\rm saturation})^2$ for smaller $\alpha$'s. Therefore the measurement of resonant part of the power spectrum is the most sensible test of these models.

In the next section we will investigate generalizations of this resonant model by letting the additional operators in the Effective Theory that we neglected so far become relevant. We will try to see if they can lead to a significant level of non-Gaussianity with a larger signal to noise with respect to the 2-point function. The answer will be that this is not the case.

%%%%%%%%%%%%%%%%%%%%%%%%%%%%%%%%%%%%%%%
%%%%%%%%%%%%%%%%%%%%%%%%%%%%%%%%%%%%%%%
%%%%%%%%%%%%%%%%%%%%%%%%%%%%%%%%%%%%%%%

\section{Generalization}

Within the EFT of Inflation, one can easily generalize resonant models to incorporate small speed of sound or large inflaton self-interactions. It is enough to add higher order geometric operators to \eqref{effaction}, which after expansion in terms of $\pi$ (restoring gauge invariance) result in new interactions. The coupling coefficients in these new interactions will naturally have small oscillating components which generate resonant non-Gaussianities similar to ones studied before. However, there is an additional way in which resonant effects can be important. So far we have simply considered the effect of an oscillating coupling on a standard Bunch-Davies wavefunction as the coupling was already of order $\epsosc$. Here this correction to the Bunch-Davies wavefunction will need to be considered as the interactions are not necessarily suppressed by $\epsosc$ anymore.

% correction to the wavefunction due to resonances will
%As explained in \S\ref{simplest} (also see~\cite{Flauger2}) the amplification of non-Gaussianities happens as a result of a resonance between the frequency of the interacting modes and the time dependent couplings at around $k_{physical}\sim\omega$. Therefore for the resonant enhancement to occur the new interactions have to have an oscillating component which pumps energy into the modes as they emerge from sub-horizon scales. There are two possibilities:
%\begin{enumerate}
%\item to incorporate oscillations in the coefficients of newly added interactions, or,
%\item to invoke the corrections to the Bunch-Davies wavefunction due to the oscillating background.
%\end{enumerate}

We will show in the following that the conclusion of the last section is generically true in these two additional cases, namely, the higher order correlators remain smaller than the modificatons of the 2-point function as long as the effective theory is valid and natural. However there is an extra enhancement of the bi-spectrum in the large folded limit in the second case above. As we will see, this will make the signal-to-noise originating from non-Gaussianities larger than the one in the 2-point function only in a very small marginal region of parameter space where the bound on $\alpha$ is saturated.
%%%%%%%%%%%%%%%%%%%%%%%%%%%%%%%%%%%%%%%
%%%%%%%%%%%%%%%%%%%%%%%%%%%%%%%%%%%%%%%
%%%%%%%%%%%%%%%%%%%%%%%%%%%%%%%%%%%%%%%

\subsection{\label{coupling} Oscillating Couplings}

In this subsection we show that the oscillating coefficients of self-interactions of $\pi$ (if present) are tied by quantum loops to the modification of the power spectrum in such a way that the contribution to the signal to noise from higher correlators is always suppressed, at least if we do not consider fine-tuned theories.

Let us consider the simplest scenario as a benchmark. Substituting \eqref{eosc} into \eqref{lineq} the linear equation of motion for $\pi$ schematically takes the form
\beq
\label{lineq2}
\ddot\pi+3H\left(1+\epsosc \alpha\cos( \omega t)\right)\dot\pi-\partial^2 \pi=0\,.
\eeq
This modification of the equation of motion excites a negative frequency part in the mode-functions \cite{Flauger1,Flauger2} (this will also be discussed in more details in \S\ref{mode_function}) which leads to
\beq
\label{power}
\delta\Expect{\R^2}\sim \epsosc\alpha^{1/2}\Rtwo\,.
\eeq

Now consider adding the following cubic interaction
\beq
\label{cubic}
\left(1+\lambda\cos(\omega t)\right) \frac{\dot\pi_c^3}{\Lambda^2}\,,
\eeq
where $\Lambda$ is the high energy cutoff of the theory at which the interaction $\dot\pi^3$ becomes strongly coupled. If we let loop corrections run until the strong coupling scale $\Lambda$ (a necessary condition if we do not give an explicit UV completion cutting off the loops and making them convergent), the contribution of this coupling to the 1-loop renormalization of the kinetic term is of order
\beq
\left(1+\lambda\cos(\omega t)\right)^2\dot\pi_c^2\, .
\eeq
Neglecting numerical factors, now the linear equation of motion becomes
\beq
\ddot\pi+3H\left(1+(\epsosc+\lambda) \alpha \sin(\omega t)\right)\dot\pi-\partial^2\pi=0\,.
\eeq
We see that the loop correction effectively renormalized $\epsosc$. In order for this renormalization to be at most of order one, so that the theory is technically natural, we need to impose
%Comparing with \eqref{lineq2} reveals that in order not to renormalize relevantly $\epsosc$, $\lambda$ is constrained to be
\beq
\lambda\lsim\epsosc \,.
\eeq
The level of resonant non-Gaussianity produced by the second term of \eqref{cubic} is easy to estimate following sec.~\ref{simplest}. We obtain 
\beq
\frac{\Expect{\R^3}}{\ \ \langle\R^2\rangle^{3/2}}\lsim \epsosc \alpha^{1/2}\left(\frac{\omega}{\Lambda}\right)^2\,,
\eeq
where we have inserted a factor of $\alpha^{1/2}$ coming from the saddle point approximation. This is always smaller than \eqref{power} within the range of validity of the effective theory $\omega\ll \Lambda$. 
%In other words making \eqref{cubic} larger, which is achieved by decreasing the cutoff $\Lambda$, does not result in a larger resonant non-Gaussianity because at the same time the acceptable range of $\alpha$ shrinks proportionally to smaller values.  
In the computation of  the non-Gaussianity at leading order in $\epsosc$, we should consistently also include the contribution from the perturbed wavefunction applied to a non-oscillatory coupling. We will describe this contribution (which turns out to be the leading one) in the next section. Here we concentrate only on the non-Gaussianity resulting from considering the oscillatory coupling on the unperturbed wavefunctions.

The arguments above can be easily generalized to any other interaction. For instance the resonant 4-point correlator in the presence of a large quartic interaction $\dot\pi^4$ as in \cite{Senatore:2010jy} is constrained to
\beq
\frac{\Expect{\R^4}}{\ \langle \R^2\rangle^2}\lsim \epsosc \alpha^{1/2}\left(\frac{\omega}{\Lambda}\right)^4\,.
\eeq
We will complete the study of this case in sec.~\ref{sec:four-point}.
Thus, we conclude that by adding an oscillating part to the couplings one cannot escape the observability bound of previous section.

Notice that in doing these estimates we have always used the cutoff of the theory and the canonically normalized $\pi$. By following the same steps as in~\cite{Senatore:2010jy}, we can easily see that even in the case of small speed of sound we do not obtain an enhancement of the non-Gaussianities because for any $c_s$ the non-gravitational sector of the Lagrangian can be made effectively Lorentz invariant by rescaling $\vec x=c_s \vec{\tilde x}$ and defining $\pi_c=(-2\mpl^2\dot H c_s)^{1/2}\pi$ and reabsoring the remaining factors of $c_s$ in a rescaled cutoff $\Lambda$~\cite{Senatore:2010jy}. Notice that now there is a stronger upper bound on $\alpha$ coming from imposing $\omega$ to be within the regime of validity of the effective theory. In this case the UV cutoff is given by \cite{Cheung:2007st}
\beq
\Lambda^4 \simeq 16\pi^2\mpl^2\dot H c_s^5\,,
\eeq
which after using $\Rtwo=H^4/4\mpl^2\dot H c_s$ yields
\beq
\label{alphacs}
\alpha_{\rm saturation}(c_s)=\frac{\Lambda}{H}= c_s\left(\frac{2\pi}{\Rtwo^{1/2}}\right)^{1/2}\simeq c_s \alpha_{\rm saturation}(c_s=1)\,.
\eeq
Since the ratio of the signal to noise of the 3-point function with respect to the 2-point function goes proportionally $\alpha^2/c_s^2$, we explicitly see that taking small $c_s$ does not help to increase the relative signal to noise ratio for the 3-point function~\footnote{Later we will see that there is an enhancement of the relative signal-to-noise coming from the folded limit of the triangles by a factor of order $\sqrt{\alpha}$. So even in this case the maximum relative signal to noise decreases by reducing~$c_s$.}. 

This conclusion can also be extended to multi-field inflation. In the case when some additional light fields ($\sigma$) have an effect on the duration of inflation, there are two possible sources of large non-Gaussianity~\cite{Senatore:2010wk}. The first is due to their self-interactions. In this case our analysis of single field inflation applies equally well to correlators of $\sigma$'s. However even with negligible self-interactions, non-Gaussianity may be generated because $\sigma$ fields might affect the duration of inflation in a non-linear way. This is the second way in which non-Gaussianities can be generated: that is through the coefficient of proportionality between $\zeta$ and $\sigma^2$. If $\partial^2\zeta/\partial\sigma^2$ is oscillatory, one might hope to have oscillatory non-Gaussianities of the local kind. However, it is very unclear how one could imagine not having a comparable-size oscillatory component in the linear coefficient $\partial\zeta/\partial\sigma$, which would lead to an oscillatory component in the 2-point function that dominates the signal. 

%%%%%%%%%%%%%%%%%%%%%%%%%%%%%%%%%%%%%%%
%%%%%%%%%%%%%%%%%%%%%%%%%%%%%%%%%%%%%%%
%%%%%%%%%%%%%%%%%%%%%%%%%%%%%%%%%%%%%%%
\subsection{Resonance from Corrections to the Mode-functions -- Folded Shapes}

Another source of resonant non-gaussianity in theories with large deviation from $c_s=1$, or more generally in theories with enhanced cubic derivative self-interactions, is the correction to the Bunch-Davies wavefunction due to the oscillating background (see e.g. \eqref{lineq2}). This correction basically amounts to the addition of a small negative-frequency part to the wavefunction after the momentum matches the resonance frequency. In the original setup without large derivative self-interactions, these
corrections could be neglected because the cubic and higher order interactions generated by the expansion of $\cos\omega(t+\pi)$ are already proportional to the small amplitude of oscillations $\epsosc$, and so this effect would be of order $\epsosc^2\ll\epsosc$.

We will show now that in the presence of large interactions these contributions to the non-Gaussianity have a comparable size with the previously considered cases (i.e. $\Expect{\R^3}\sim \epsosc\alpha^{5/2}\R^4$), and thus they should be taken into account. Even more importantly, it is further enhanced in the limit of folded triangles. This is a general feature of non-Gaussianities generated through the modification of the vacuum wavefunction \cite{Kachru}, which have already been calculated for resonant models in \cite{Chen3} by means of approximating the coefficient of the negative frequency mode by a smoothed step function. Here we avoid this approximation and rely instead on the saddle point method. We recover the same scaling, namely $\Expect{\R^3}\propto \alpha^{5/2}$ away from the folded limit and $\Expect{\R^3}\propto \alpha^{7/2}$ in the folded limit. We emphasize that such an enhanced 3-point function signal can potentially escape the observability constraint of the last section when $\alpha$ is close to its upper bound.

Following \cite{Flauger2}, we will first review the derivation of the modification to the Bunch-Davies vacuum and then compute its contribution to the resonant non-Gaussianity. Later we present the results of numerical computation of the signal to noise ratio, showing that despite the enhancement in the folded limit the 3-point function remains subdominant in practically the whole relevant region of parameter space. The reader who is not interested in the technical details can refer directly to \eqref{3_point} and subsequent equations for final results.

%%%%%%%%%%%%%%%%%%%%%%%%%%%%%%%%%%%%%%%
%%%%%%%%%%%%%%%%%%%%%%%%%%%%%%%%%%%%%%%
%%%%%%%%%%%%%%%%%%%%%%%%%%%%%%%%%%%%%%%

\subsubsection{\label{mode_function} Correction to the Wavefunction}

Let us consider the effect of adding the following operators
\beq
\label{M2M3}
\frac{1}{2!}M_2^4\left(\delta g^{00}\right)^2+\frac{1}{3!}M_3^4\left(\delta g^{00}\right)^3\,,
\eeq
to the minimal action \eqref{effaction}. After restoring diff. invariance the quadratic action for $\pi$ becomes
\beq
\label{S_2_cs}
S=\int d^4x a^3\left[-\frac{\mpl^2\dot H}{c_s^2}(\dot\pi^2-c_s^2(\partial_i\pi)^2) \right]\,,
\eeq
where
\beq
c_s^2\equiv\frac{-\mpl^2\dot H}{-\mpl^2\dot H+2M_2^4}\,.
\eeq
The linear equation of motion reads
\beq
\label{lineqxcs}
\partial_x^2\pi_{\bf k}-\frac{2(1+\delta)}{x}\partial_x\pi_{\bf k}+\pi_{\bf k}=0\,,
\eeq
where we defined $x=-c_s k \tau$ and $\delta$ is now generalized to
\beq
\delta=\frac{c_s^2}{2H\dot H}\frac{d}{dt}\left(\frac{\dot H}{c_s^2}\right)\,.
\eeq
This equation is identical to the one studied before in \eqref{lineqx}. 
We assume that $\delta$ has an oscillating component
\beq
\label{delta}
\delta_{osc}=-\frac{1}{2}\epsosc\alpha\sin\omega t\,,
\eeq
as used to be the case when $c_s=1$, and it is a generalization of the definition of $\epsosc$ to the $c_s\neq 1$ case. The constant part of $\delta$ can be safely neglected as the resulting effect is as usual slow-roll suppressed.

This oscillatory component excites the Bunch-Davies vacuum \cite{Flauger1}, and the change in the usual vacuum solution to first order in $\epsosc$ can be written as a small admixture with negative frequency mode
\beq
\label{pi_x}
\pi_{\bf k}(x)=\pi_{\bf k}^{(0)}[u_+(x)+{c_k^{(-)}}(x)u_-(x)]=\pi_{\bf k}^{(0)}[(1-ix)\e^{ix}+{c_k^{(-)}}(x)(1+ix)\e^{-ix}]\,,
\eeq
with $\pi_{\bf k}^{(0)}=(\sqrt{4\mpl^2\ep c_s}k^{3/2})^{-1/2}$\,.
Substituting \eqref{pi_x} in \eqref{lineqxcs} and solving perturbatively in $\delta$, we obtain at linear level~\footnote{The explicit analytic solution can be found and the correction to the unperturbed wavefunction reads:
\bea\nonumber
&&\delta\pi=\frac{\epsosc}{\sqrt{4\mpl^2\epsilon c_s}}\frac{ 1}{4 \left(\alpha ^2+1\right) k^{3/2}}
e^{-i \eta  k} (-\eta H)^{-i \alpha } \left(\left(\alpha
   +i\right)^2+\alpha  e^{2 i  k \eta} (  k \eta +i) \left(\left(\alpha -i\right)^2 (-\eta H)^{2 i
   \alpha } E_{1-i \alpha }(2 i k \eta )\right.\right.\\ \nonumber 
   &&\left.\left. -   \left(\alpha +i\right)^2 E_{i \alpha +1}(2 i k \eta
   )\right)+(\alpha -i) (-\eta H)^{2 i \alpha } (\alpha -i \alpha   k \eta + 
   k \eta-i)-i \left(\alpha ^2+1\right)  k \eta\right)\ ,
\eea
where $E_{\nu}(x )$ represents the exponential integral function of index $\nu$.
}
\beq
\label{c_-full}
{c_k^{(-)}}(x)=\int^x dx'\;{e^{2ix'}x'^2\over (1+ix')^2}\int^{x'} {2(1+ix'')\delta\over x''^2} dx''\ .
\eeq
Clearly $\pi_{\bf k}(x)$ freezes at late times which is evident from the quickly vanishing ${d c_k^{(-)}/ dx}\sim  x$ for small $x$.  The expression above can be simplified for the resonance period which takes place well inside the horizon $x\gg 1$.
Using \eqref{delta} and noting that $\omega t = -\alpha\ln(x/k)$ plus a $k$-independent constant, one obtains
\beq
\label{c_-}
{c_k^{(-)}}(x)\simeq i\epsosc \int_\infty^xdx'\e^{2ix'}\cos(\omega t(x')) \,.
\eeq
As thoroughly discussed in \cite{Flauger1}, ${c_k^{(-)}}(x)$ starts from zero at early times ($x\to \infty$), jumps at resonance period $x\sim\alpha/2$ and stays constant afterwards. Both the step-like behavior and the small oscillations on top of that are important to obtain the resonant behavior and the additional enhancement in the folded limit. 

Expression (\ref{c_-}) gives a good approximation for $x$ not much smaller than one and so it covers  the resonance period $x\gtrsim 1$. Therefore we will use this simplified expression in the calculation below. Let us note that (\ref{c_-}) will make some of the terms, for example the contribution to the 3-point function from the vertex $\dot{\pi}(\partial_i\pi)^2$, IR divergent.
This is a fake divergence which will disappear once one uses the full expression (\ref{c_-full}) which is different from (\ref{c_-}) for small $x$. Equivalently, to get the correct result at leading order  one can simply use~(\ref{c_-}) throughout and simply disregard the contribution from the small $x$ region.

The value of $c_{k}^{(-)}$ at late times, when $x\to0$  can still be well approximated with the saddle point approximation for (\ref{c_-}), since the phase is stationary at $x\gg1$. We obtain
%Finally, we can use the saddle point approximation to evaluate $c_{k}^{(-)}$ at late times to get
\beq
\label{c_-_0}
{c_k^{(-)}}(0)\simeq -i\frac{\sqrt{2\pi}}{4}\epsosc\alpha^{1/2}\e^{i\alpha\ln 2k/k_\star}\,,
\eeq
and the power spectrum can at this point be easily derived to obtain (\ref{2pt}).
%up to an unimportant $k$-independent phase which will be ignored throughout the discussion.

%%%%%%%%%%%%%%%%%%%%%%%%%%%%%%%%%%%%
%%%%%%%%%%%%%%%%%%%%%%%%%%%%%%%%%%%%%%%
%%%%%%%%%%%%%%%%%%%%%%%%%%%%%%%%%%%%%%%

\subsubsection{\label{3pt}3-point function}

After expanding~\eqref{M2M3} in terms of $\pi$ the cubic action reads as follows
\beq
\label{S_3_cs}
S_3=\int d^4x a^3\mpl^2\dot H\left[\lambda_{\dot{\pi}^3}c_s^{-4}\dot\pi^3+(c_s^{-2}-1)a^{-2}\dot\pi(\partial_i\pi)^2\right]\, \\ \nonumber
\text{with}\quad\lambda_{\dot{\pi}^3}\mpl^2\dot H c_s^{-4}\equiv-\mpl^2\dot H(c_s^{-2}-1)-\frac{4}{3}M_3^4\ ,
\eeq
where as discussed in \cite{Cheung:2007st} $M_3$ is a free parameter. It is technically natural to choose $M_3^4\sim M_2^4/c_s^2\sim -\mpl^2\dot H/c_s^4$, so that $\dot\pi^3$ gives the same level of non-Gaussianity as $\dot\pi(\partial_i\pi)^2$. This choice corresponds to $\lambda_{\dot\pi^3}\sim 1$. As mentioned earlier, in the presence of large cubic self-interactions, modification of the wavefunction can generate large non-Gaussianities. At lowest order it suffices to use the negative frequency term, instead of the positive-frequency one, for one of the fields in the calculation of the 3-point function using first order perturbation theory. For instance from $\dot\pi^3$ we get
\beq
\label{pi_3point}
\Expect{\pi^3}_{\dot\pi^3}&=&(-i)\lambda_{\dot{\pi}^3}\mpl^2\dot H c_s^{-4}\left[\prod_{i=1}^3\frac{1}{4\mpl^2\ep c_sk_i^3}\right]\frac{-1}{H}\int_{-\infty}^0\frac{d\tau}{\tau}({{c^{(-)}}^*u_-^*}_{k_1})'{u_+^*}_{k_2}'{u_+^*}_{k_3}' \nonumber\\
&&~~~~~~~~~~~~~~~~~~~~~~~~~~~~\qquad+\text{perm.}+\rm{c.c.}+\text{other choices of $c^{(-)}$}
\,,
\eeq
where $u_+$ and $u_-$ are defined in \eqref{pi_x} as the positive and negative frequency solutions and we have dropped their arguments to avoid notational clutter. Prime denotes derivative with respect to the conformal time $\tau$. It can be straightforwardly checked that substituting the negative modes in the final wavefunctions gives a negligible effect because there is no resonance enhancement. 

Because of the oscillating behavior of $c_k^{(-)}(x)$ (see \eqref{c_-}) the integral is very similar to the one discussed in the previous subsection. This explains the scaling $\Expect{\R^3}\sim \epsosc\alpha^{5/2}\Rtwo^2$ away from the folded limit. The further enhancement in the folded limit is due to 
the fact that in this limit the phase of the integral is almost constant and $c_-$ can be taken to be a step function that rises from 0 to $\epsosc\alpha^{1/2}$ at resonance frequency $k\tau\sim- \alpha$. This effectively enhances the time integral by a factor  of $\alpha$. %The enhancement of the folded limit is also expected from the step-like behavior of $c_k^{(-)}$ \cite{Chen3}, because close to the limit the phase of the integral changes very slowly and $c_k^{(-)}$ can be approximated to be turned on sharply which gives a large contribution. In the opposite limit $c_k^{(-)}$ can be thought of as being turned on adiabatically.

Now we will show that since the resonance happens well inside the horizon, when $|k\tau|\gg 1$,  in the integral
\beq
I_{\dot{\pi}^3}=\int_{-\infty}^0\frac{d\tau}{\tau}({{c^{(-)}}^*u_-^*}_{k_1})'{u_+^*}_{k_2}'{u_+^*}_{k_3}'\
\eeq
we can keep only the terms with highest power of $\tau$. Let us use the variables
\beq
\label{x_i}
x_i=-c_sk_i\tau\,,\quad \text{and}\quad y_i=1-\frac{1}{k_i}\sum_{j\neq i}k_j\,,
\eeq
which are useful because $y\rightarrow 0^-$ corresponds to the folded limit. 
$I_{\dot{\pi}^3}$ can be written as
\beq
\label{I}
I_{\dot{\pi}^3}(y_1)=\frac{c_s^3k_2^2k_3^2}{k_1}\left(I_{\rm I}{}^{(2)}(y_1)+I_{\rm II}{}^{(1)}(y_1)-iI_{\rm II}{}^{(2)}(y_1)\right)\,,
\eeq
where we introduced
\beq
\label{I_1}
I_{\rm I}^{(n)}(y_1)=\int_0^\infty dx_1 x_1^n \e^{ix_1y_1} {c^{(-)}} ^*(x_1)\,,\qquad\quad\quad
\label{I_2}
I_{\rm II}^{(n)}(y_1)=\int_0^\infty dx_1 x_1^n \e^{ix_1y_1} \frac{d}{dx_1}{c^{(-)}} ^*(x_1)\,. \nonumber \\ 
\eeq
We only know an integral representation for ${c^{(-)}}(x_1)$ and therefore it is convenient to integrate $I_I^{(n)}$ by parts to obtain
\beq
\label{integralI1}
I_{\rm I}^{(n)}(y_1)=(-i)^n\frac{\partial^n}{\partial y_1^n}\left[\left.\frac{\e^{ix_1y_1}-1}{iy_1}{c^{(-)}}^*(x_1)\right\arrowvert_0^\infty-\frac{1}{iy_1}\int_0^\infty dx_1 (\e^{ix_1y_1}-1) \frac{d}{dx_1}{c^{(-)}}^*(x_1)\right]\,. \ \ \
\eeq
Since ${c^{(-)}}$ is zero at early times, ${c^{(-)}}(x\to\infty)=0$, the first term in (\ref{integralI1}) vanishes.
With help of the explicit form of ${dc_{k}^{(-)}/ dx}$ which can be extracted from (\ref{c_-})
the remaining integrals can be calculated using the saddle point approximation.
%\beq
%\label{saddle}
%\int_0^{\infty(1-i\varepsilon)} e^{-ix}e^{i\alpha \ln x}\simeq \sqrt{2\pi \alpha} e^{i\alpha(\ln\alpha-1)}\left(1+{\mathcal O}\left({1\over \alpha}\right)\right)
%\eeq
%assuming a large and positive $\alpha$. When $\alpha$ is large and negative the corresponding integral is suppressed by $e^{-\alpha}$.

The leading order in $\alpha$ contribution comes from the highest derivative which acts on $y_1$ in the argument of the exponent:
\beq
\label{apprxII}
\left[i{\epsosc \alpha^{1/2}\sqrt{\pi}\over \sqrt{2}}e^{-i\alpha \ln(k_1/k_\star)}\right]^{-1}I_{\rm II}^{(n)}={\mathcal I}_{\rm II}^{(n)}\equiv(-i)^n{\partial^n~\over \partial y_1^n}{e^{-i\alpha\ln(2-y_1)}\over (2-y_1)} \simeq {\alpha^n e^{-i\alpha\ln(2-y_1)}\over (2-y_1)^{n+1}}\,, \nonumber \\
\eeq
and similarly
\beq
\label{tricky}\nonumber
\left[i{\epsosc\alpha^{1/2}\sqrt{\pi}\over \sqrt{2}}e^{-i\alpha \ln(k_1/ k_\star)}\right]^{-1}I_{\rm I}^{(n)}={\mathcal I}_{\rm I}^{(n)} \equiv (-i)^n{\partial^{n}~\over \partial y_1^n}\left.{ie^{-i\alpha\ln(2-y)}\over y_1(2-y)}\right|_{y=0}^{y=y_1} \simeq {i\alpha^n e^{-i\alpha\ln(2-y_1)}\over y_1(2-y_1)^{n+1}}\ .\\
\eeq
The approximation (\ref{apprxII}) is valid for any $y_1$ while the last passage of (\ref{tricky}) is only valid for sufficiently large $|y_1|\geq 2 n \alpha^{-1}$ far away from the folded limit. In the folded limit $y_1\rightarrow 0$ we can use the expression for ${\mathcal I}_{\rm I}^{(n)}$ as it approaches $y_1\rightarrow 0^-$
\beq
\label{rightap}
{\mathcal I}_{\rm I}^{(n)} \quad\rightarrow\quad  -{\alpha^{n+1} e^{-i\alpha\ln 2}\left(1+{i\alpha y_1 (n+1)\over 2(n+2)} - {\alpha^2 y_1^2(n+1)\over 8 (n+3)}+\dots\right)\over 2^{n+2}(n+1)}\ .
\eeq
In Figure \ref{Figy1} we compare the explicit expression for ${\mathcal I}_{\rm I}^{(2)}$ with the approximations (\ref{tricky},\ref{rightap}). For large $\alpha$ the approximate expressions give a good agreement for both $y_1\rightarrow 0^-$ and $y_1\lesssim-1/\alpha$ regions. In the region $y_1\sim -1/\alpha$ the exact expression for ${\mathcal I}_{\rm I}^{(n)}$ oscillates with a rapidly changing amplitude and both approximations break down~\footnote{This means that for $y_1\sim- 1/\alpha$, which corresponds to a quasi folded limit, we should use the exact expression, but at this point the formulae become bulky and we do not compute them explicitly. However from the plot we see that this region corresponds to simple interpolation between the two extreme regions where the expressions are simple, without any important change.}.

%%%%%%%%%%%%%%%%%%%%%%%%%%%%%%%%%%%%%%%%%%%%%%%%%%%%%%%%%%%%%%%%%%%%%%%%%%%%
\begin{figure}[h]
\begin{center}
\includegraphics[width=2.8in]{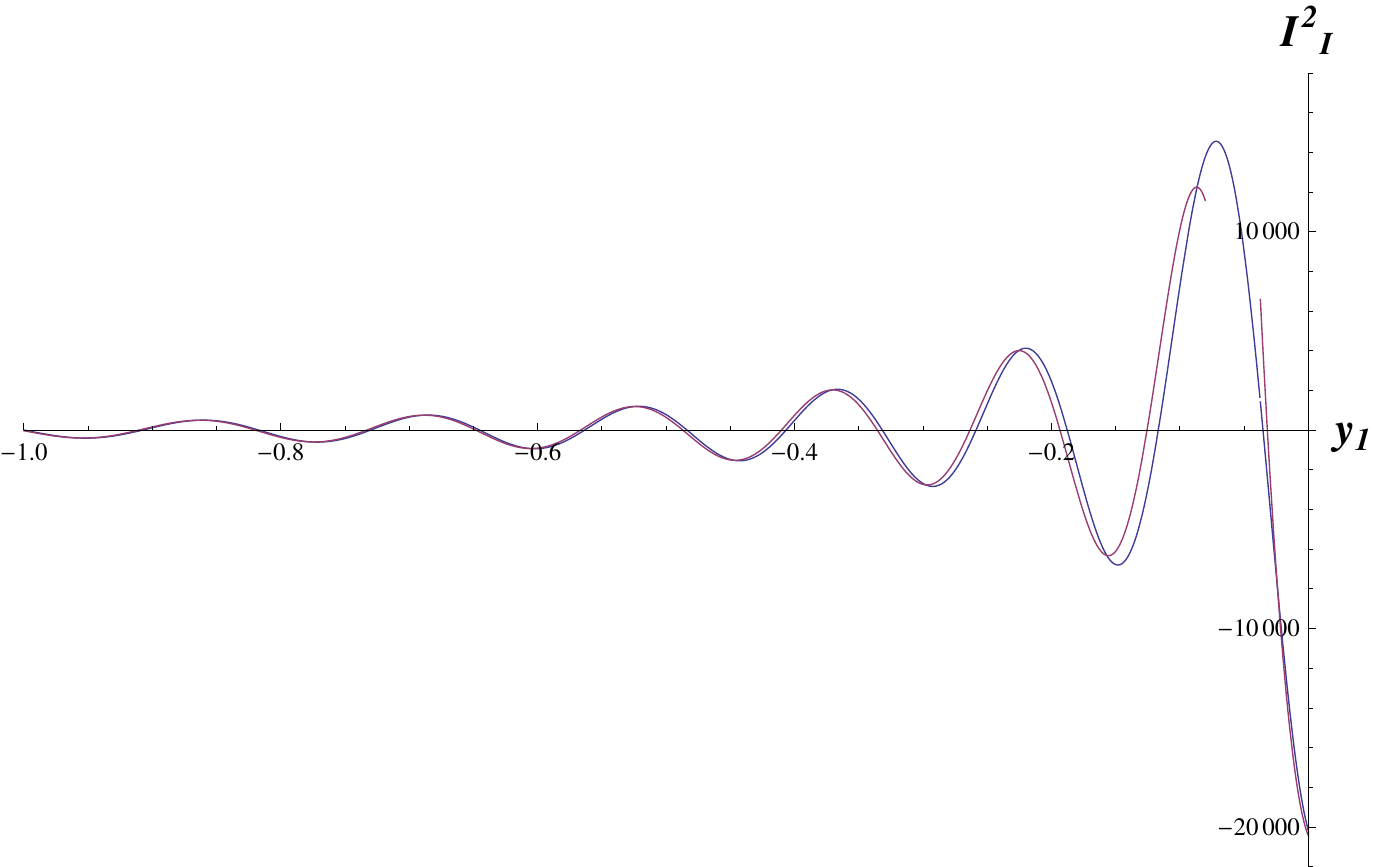}
\caption[]{ Explicit expression for  ${\mathcal I}_{\rm I}^{(2)}$ for $\alpha=100$ and the approximations (\ref{tricky},\ref{rightap}) for large and small $y_1$ correspondingly.  }
\label{Figy1}
\end{center}
\end{figure}
%%%%%%%%%%%%%%%%%%%%%%%%%%%%%%%%%%%%%%%%%%%%%%%%%%%%%%%%%%%%%%%%%%%%%%%%%%%%%

Since $I_{\rm I,II}^{(n)}\sim \alpha^{n-1/2}$ we conclude that only ${\mathcal I}^{(n)}$ with the highest $n$ are important in (\ref{I}), confirming that we can keep only the terms with the highest powers of  $\tau$.
Substituting the expressions for $I^{(2)}_{\rm I}$ and $I^{(2)}_{\rm II}$, summing over permutations and using $\R\simeq -H\pi$ we finally find the correlator of three $\R$'s away from the folded limit
\beq
\label{3_point}
\!\!\!\!\!\!\Expect{\R^3}_{\dot{\pi}^3} \simeq\!-3\sqrt{\frac{\pi}{2}}\lambda_{\dot{\pi}^3}c_s^{-2}\Delta_\zeta^2\epsosc\alpha^{5/2}\left(\sum_i\frac{1}{y_i}-3\right)\frac{1}{k_1k_2k_3K^3}\sin(\alpha\ln K/k_\star) \,,
\eeq
 with $K=k_1+k_2+k_3$ and we have introduced $\Delta_\zeta= H^2/(4\ep c_s\mpl^2)$. In the folded limit $k_1\rightarrow k_2+k_3$, we have
\beq
\label{3p_fold}
\!\!\!\!\!\!\!\!\!\!\!\!\Expect{\R^3}^{\it folded}_{\dot{\pi}^3}\simeq \frac{1}{32}\lambda_{\dot{\pi}^3}c_s^{-2}\Delta_\zeta^2\sqrt{2\pi}\epsosc\alpha^{7/2}{k_1^4k_2k_3}\cos(\alpha\ln 2 k_1/k_\star)\,.
\eeq
As anticipated, we see that away from the folded limit we have the same scaling as before $\sim\alpha^{5/2}$, while at the folded limit we have $\sim\alpha^{7/2}$.  This expression holds approximately for a region in $y_1\sim[-1/\alpha,0]$.

%%%%%%%%%%%%%%%%%%%%%%%%%%%%%%%%%%%%%%%%%%%%%%%%%%%%%%%%%%%%%%%%%%%%%%%%%%%%%%
\begin{figure}[t]
\begin{center}
\includegraphics[angle=0,width=80mm]{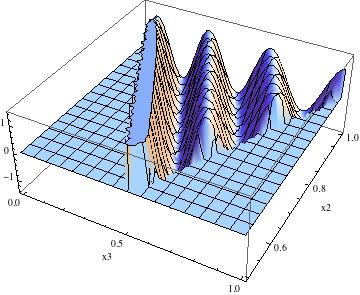}
\caption{ The shape of non-gaussian signal (\protect \ref{3_point}) induced by vertex  $\dot\pi^3 $ plotted for $\alpha=50$ away from the folded limit as a function of $x_i=k_i/k_1$.} %\eqref{3_point}
\label{Fig: pidot}
\end{center}
\end{figure}
%%%%%%%%%%%%%%%%%%%%%%%%%%%%%%%%%%%%%%%%%%%%%%%%%%%%%%%%%%%%%%%%%%%%%%%%%%%%%
%%%%%%%%%%%%%%%%%%%%%%%%%%%%%%%%%%%%%%%%%%%%%%%%%%%%%%%%%%%%%%%%%%%
\begin{figure}[htbp]
\begin{center}
\includegraphics[angle=0,width=80mm]{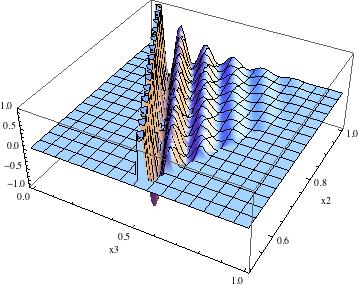}
\caption{ The shape of non-gaussian signal for the 3-point function generated by the vertex $\dot\pi(\partial_i\pi)^2 $ (\protect \ref{pidotgradpi}) plotted for $\alpha=100$ away from the folded limit. \label{Fig: pidotgradpi}}
\end{center}
\end{figure}
%%%%%%%%%%%%%%%%%%%%%%%%%%%%%%%%%%%%%%%%%%%%%%%%%%%%%%%%%%%%%%%%%%%%%

The computation of the contribution to the 3-point function due to the vertex $\dot\pi(\partial_i\pi)^2$ is similar
\beq
\label{pi_3point}
\Expect{\pi^3}_{\dot{\pi}(\partial_i\pi)^2}&=&(-i)\mpl^2\dot H (c_s^{-2}-1)\left[\prod_{i=1}^3\frac{1}{4\mpl^2\ep c_sk_i^3}\right]\frac{-1}{H} I'\quad+\text{perm.}+\rm{c.c.}+\text{other choices of $c^{(-)}$} \nonumber \\
\eeq
\beq
I'&=&c_s\frac{k_2k_3({\bf k}_2\cdot{\bf k}_3)}{k_1}(I^{(2)}_{\rm I}(y_1)-iI^{(2)}_{\rm II}(y_1))\nonumber\\
&&-c_s\frac{k_1^2k_3({\bf k}_2\cdot{\bf k}_3)}{k_2^2}I_{\rm I}^{(2)}(y_2)-c_s\frac{k_1^2k_2({\bf k}_2\cdot{\bf k}_3)}{k_3^2}I^{(2)}_{\rm I}(y_3)\, .
\eeq
Here we dropped the subleading $I^{(n)}_{\rm I,II}$ with $n<2$. %AD one of these terms I^{-1} is divergent!
Now it is straightforward to get the contribution to the 3-point function
\beq
\label{pidotgradpi}
\Expect{\R^3}_{\dot{\pi}(\partial_i\pi)^2}&\simeq&- \sqrt{\frac{\pi}{2}}(c_s^{-2}-1)\Delta_\zeta^2\epsosc\alpha^{5/2}\frac{1}{k_1^2k_2^2k_3^2K^3}\times  \\
&\times &\left(\left(\sum_i\frac{k_i^3}{y_i}-\sum_{i\neq j}\frac{k_ik_j^2}{y_i}\right)-\frac{1}{2}\left(\sum_ik_i^3-\sum_{i\neq j}k_ik_j^2\right)\left(1+\sum_i\frac{1}{y_i}\right)\right)\sin(\alpha\ln K/k_\star)\,, \nonumber
\eeq
%\beq
%\Expect{\R^3}&\simeq&\left(\frac{1}{c_s^2}-1\right)\R^4\sqrt{2\pi}\delta_0\alpha^{3/2}\frac{1}{k_1^2k_2^2k_3^2K^3}\\
%&&\times\left((\sum_i\frac{k_i^3}{y_i}-\sum_{i\neq j}\frac{k_ik_j^2}{y_i})-\frac{1}{2}(\sum_ik_i^3-\sum_{i\neq j}k_ik_j^2)(1+\sum_i\frac{1}{y_i})\right)\sin(\alpha\ln K)\,, \nonumber
%\eeq
away from the folded limit, and
\beq
\Expect{\R^3}^{\it folded}_{\dot{\pi}(\partial_i\pi)^2}\simeq \frac{1}{32}(c_s^{-2}-1)\Delta_\zeta^2\sqrt{2\pi}\epsosc\alpha^{7/2}\frac{1} {k_1^4 k_2 k_3}\cos(\alpha\ln 2 k_1/k_\star)\,,
\eeq
%\beq
%\Expect{\R^3}_{folded}\simeq\left(\frac{1}{c_s^2}-1\right)\R^4\frac{\sqrt{2\pi}\delta_0\alpha^{5/2}}{48}\frac{1}{k_1^4k_2k_3}\cos(\alpha\ln 2k_1)\,,
%\eeq
in the folded limit $k_1\rightarrow k_2+k_3$. 

For completeness, we also give the shape induced by the oscillatory couplings on the unperturbed wavefunction. This is given by \cite{Chen3}
\beq
\label{3ptcoupling}
\Expect{\R^3}\sim c_s^{-2}\Delta_\zeta^2\sqrt{2\pi} \epsosc\alpha^{5/2}\left[\frac{\tilde\lambda_{{\dot\pi}^3}}{k_1k_2k_3K^3}+\frac{\tilde\lambda_{\dot\pi{(\partial_i\pi)}^2}}{k_1^2k_2^2k_3^2K^3} \left(\sum_ik_i^3-\sum_{i\neq j}k_ik_j^2\right)\right]\sin(\alpha\ln K/k_\star)\,, \nonumber \\
\eeq
where $\tilde\lambda_{\dot\pi^3},\tilde\lambda_{\dot\pi(\partial_i\pi)^2}\sim 1$, correspond  to $\dot\pi^3$ and $\dot\pi(\partial_i\pi)^2$ interactions, respectively. As described in the former subsection, we see that this contribution is also subleading in the sense of signal to noise ratio with respect to the 2-point function.
 
%%%%%%%%%%%%%%%%%%%%%%%%%%%%%%%%%%%%%%%%%%%%
%%%%%%%%%%%%%%%%%%%%%%%%%%%%%%%%%%%%%%%%%%%%
%%%%%%%%%%%%%%%%%%%%%%%%%%%%%%%%%%%%%%%%%%%%

\subsubsection{Upper Bound and Observability for the Folded Shapes}

The shapes of the 3-point functions we computed are very different from the standard ones that are analyzed. Indeed, because of the oscillations in $k$-space, in App.~\ref{app:cosines} we show that for reasonable high $\alpha$'s they become orthogonal to the standard shapes. This means that current bounds on non-Gaussianities do not relevantly  constrain these shapes. However, we are now going to see that the leading signal-to-noise for these models is in the 2-point function.

Because of the enhancement in the folded limit, the signal-to-noise ratio can be enhanced with respect to the naive one. Indeed the naive inequality $\Expect{\R^3}/\Rtwo^{3/2}\leq \delta\Expect{\R^2}/\Rtwo$, is not satisfied  for configurations close to the folded limit, and so one might wonder if the enhanced signal-to-noise in the 3-point function coming from those  configurations can potentially make the 3-point function the leading observable. The shape in the folded limit was given by eq.~\eqref{3_point}. This shape is valid for a region of parameters space from $y\sim 0^-$ to $y\sim 1/\alpha$. The 3-point function is enhanced in this region by  a factor of $\alpha$ so that the signal-to-noise gets enhanced just proportionally to $\sqrt{\alpha}$.

%%%%%%%%%%%%%%%%%%%%%%%%%%%%%%%%%%%%%%%%%%%%%%%%%%
\begin{figure}[t]
\begin{center}
\includegraphics[angle=0,width=70mm]{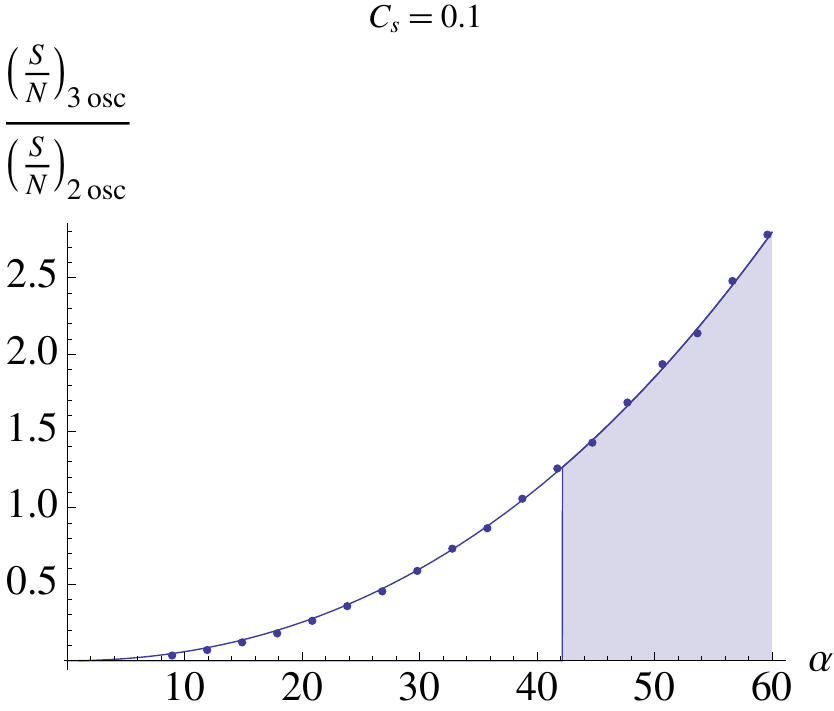}
\includegraphics[angle=0,width=70mm]{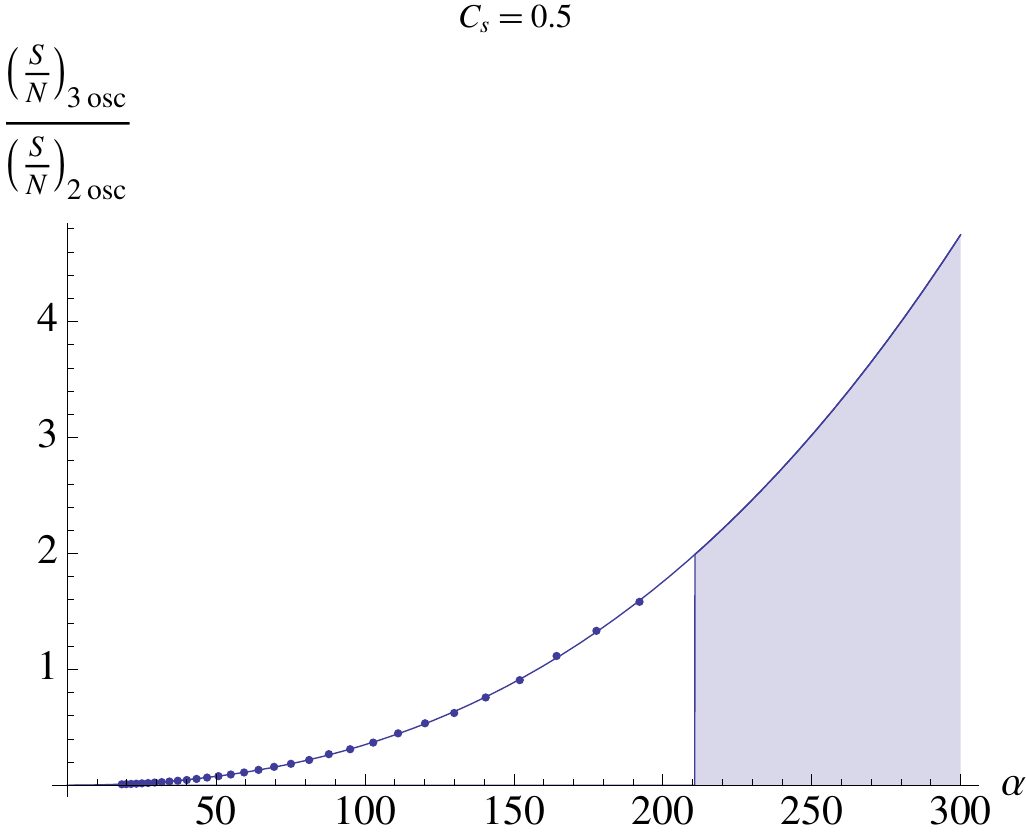}
\caption[]{\label{Fig: S_N} Ratio of the signal to noise ratio for the 3-point function generated by vertex $\dot\pi^3$ (\ref{3_point}, \ref{3p_fold}) to the signal to noise ratio of the resonant correction to the 2-point function (\protect \ref{2pt}) plotted as a function of $\alpha$ for $c_s=0.5$ on the left, and $c_s=0.1$ on the right. The shaded region corresponds to the values of $\alpha$ where the effective theory breaks down, and the allowed values of $\alpha$ should be well within the white region. We see that the region where the signal to noise from the 3-point function is bigger than the signal to noise from the 2-point function is irrelevantly small and corresponds to when the theory is not under parametric control.}
\end{center}
\end{figure}

%%%%%%%%%%%%%%%%%%%%%%%%%%%%%%%%%%%%%%%%%%%%%%%%%%%%%%%%%%%%%%%%%%%%%%%%%%%%%%%

In Fig.~\ref{Fig: S_N} we plot the ratio of the signal to noise for the 3-point function versus the one for the 2-point function. We see that there is a very small region in which this ratio is larger than one.  This happens only when $\alpha$ becomes extremely close to its upper bound. Notice that in order to have a controlled perturbative expansion the parameter $\alpha$ needs to be much smaller than that. We therefore conclude that  the leading signal to noise comes from the 2-point function.

%We saw that the large interactions reduce the UV cutoff and the upper-bound of $\alpha$. Therefore to have the largest possible $\alpha$ we should allow only modest interactions or equivalently take $c_s\sim 1$ (see eq. \eqref{alphacs}). It follows from a numerical analysis (similar to the one done in Appendix A) that the folded limit enhancement can improve the signal to noise ratio of the folded shapes compared to the power spectrum modifications by a factor of $\sim 2$. The result are presented in figure \ref{Fig: S_N}. Considering  that many coefficients are subject to an order-one change, it might be interesting to check the folded shapes against the CMB data.

%%%%%%%%%%%%%%%%%%%%%%%%%%%%%%%%%%%%%%%%%%%%%%%%%%%%%%%%%%%%%%%%%
%%%%%%%%%%%%%%%%%%%%%%%%%%%%%%%%%%%%%%%%%%%%%%%%%%%%%%%%%%%%%%%%

\subsubsection{4-point function\label{sec:four-point}}

In~\cite{Senatore:2010jy}, it was shown that there are technically natural models of single clock inflation where the 4-point function has a larger signal to noise than the 3-point function. For models not very close to de Sitter space, the 4-point function is induced by only one operator $M_4^4\left(1+\mu_0\cos(\omega t)\right) \dot\pi^4$. Following the steps of~\cite{Senatore:2010jy}, we can rescale the spatial coordinates and canonically normalize $\pi$ to show that this term takes the form
\be
\int d^4\tilde x dt\,  a^3\; \left(1+\mu_0\cos(\omega t)\right)\frac{\dot\pi_c^4}{\Lambda_{U,4}^4}\ ,
\ee
where $\pi_c=(-2\dot H\mpl^2 c_s)^{1/2}\pi$, $\vec{\tilde x}=\vec x/c_s$, and
\be
\Lambda_{U,4}^4\sim \frac{\left(\dot H\mpl^2\right)^2}{M^4_4 c_s}\ .
\ee
Let us impose that the signal induce by the 4-point function is bigger than the one from the 3-point function. This is achieved by imposing
\be
\left.\frac{{\cal L}_4}{{\cal L}_3}\right|_{E\sim \omega}\gtrsim 1 \quad\Rightarrow\quad \Lambda_{U,4}^4\lesssim\omega^2 \Lambda_{U,3}^2=\alpha^2 H^2\Lambda_{U,3}^2\quad\Rightarrow\quad M_4^4\gtrsim \frac{\left(\dot H\mpl^2\right)^{3/2}}{\omega^2 c_s^{7/2}}\ .
\ee 
In~\cite{Senatore:2010jy} it is shown that it is technically natural for $c_s$ to be order one or smaller. The result above further implies that the maximum $\alpha$ that we can have is (compare to (\ref{alphacs}))
\be
\alpha_{\rm saturation}\sim \frac{\Lambda_{U,4}}{H} \quad\Rightarrow\quad \alpha_{\rm saturation}\sim \frac{c_s}{\langle\zeta^2\rangle^{1/4}}\ .
\ee
If we now turn to the coefficient of the oscillating term $\mu_0$, it should not renormalize the oscillating kinetic term by a large amount, implying the constraint
\be
\mu_0\lesssim \epsosc\ .
\ee

The leading contribution to the four point function consists of two parts. First, there is a contribution due to the correction to the Bunch-Davies wave-function which is $\mu_0$ independent
\beq
\label{pi_4point}
\Expect{\pi^4}_{\dot{\pi}^4}&=&(-i)M_4^4\left[\prod_{i=1}^4\frac{1}{4\mpl^2\ep c_sk_i^3}\right]\int_{-\infty}^0d\tau({{c^{(-)}}^*u_-^*}_{k_1})'{u_+^*}_{k_2}'{u_+^*}_{k_3}'{u_+^*}_{k_4}' \nonumber\\
&&~~~~~~~~~~~~~~~~~~~~~~~~~~~~\qquad+\text{perm.}+\rm{c.c.}\,
\eeq
Using the variables
\beq
\label{x_i}
x_i=-c_sk_i\tau\,,\quad \text{and}\quad y_i=1-\frac{1}{k_i}\sum_{j\neq i}k_j\,,
\eeq
the contribution above can be rewritten as
\beq
\Expect{\pi^4}_{\dot{\pi}^4}&=&(-i)M_4^4\left[\prod_{i=1}^4\frac{1}{4\mpl^2\ep c_sk_i^3}\right] \frac{c_s^3k_2^2k_3^2k_4^2}{k_1^3}\left(I^4_{\rm I}(y_1)-iI_{\rm II}^4(y_1)\right)\nonumber\\
&&~~~~~~~~~~~~~~~~~~~~~~~~~~~~\qquad+\text{perm.}+\rm{c.c.}\,,
\eeq
where we neglected the subleading contribution of $I_{\rm I,II}^{(n)}$ with $n<4$.
With the help of (\ref{apprxII}) and~(\ref{tricky}) and using $\R\simeq -H\pi$, we obtain
\beq
\label{4_point}
\!\!\!\!\!\!\Expect{\R^4}_{\dot{\pi}^4}=6 \left[-{M_4^4\over \mpl^2 \dot{H}}\right] c_s^2\Delta_\zeta^3\sqrt{2\pi}\epsosc\alpha^{9/2}\left(\sum_i\frac{1}{y_i}-4\right)\frac{1}{k_1k_2k_3k_4K^5}\sin(\alpha\ln K/k_\star) \,,
\eeq
away from the folded limit, and
\beq
\!\!\!\!\!\!\!\!\!\!\!\!\Expect{\R^4}^{\it folded}_{\dot{\pi}^4}=\left[{M_4^4\over \mpl^2 \dot{H}}\right]c_s^2\Delta_\zeta^3\frac{3\sqrt{2\pi}\epsosc\alpha^{11/2}}{160}\frac{1}{k_1^6k_2k_3k_4}\cos(\alpha\ln 2k_1/k_\star)\,,
\eeq
in the folded limit $k_1=k_2+k_3+k_4$. We see that there is an enhancement proportional to $\alpha$.

 Second, there is a contribution due to the oscillating coupling $\mu_0\cos(\omega t)$ which is also easy to calculate
\beq
\label{pi4point}
\Expect{\pi^4}_{\mu_0\dot{\pi}^4}&=&(-i)M_4^4\mu_0 \left[\prod_{i=1}^4\frac{1}{4\mpl^2\ep c_sk_i^3}\right]\int_{-\infty}^0d\tau\; {u_+^*}_{k_1}'{u_+^*}_{k_2}'{u_+^*}_{k_3}'{u_+^*}_{k_4}' \cos(\omega t) \nonumber\\
&&~~~~~~~~~~~~~~~~~~~~~~~~~~~~\qquad+\rm{c.c.}\,
\eeq
After the usual change of variables $x=-c_s K\tau$ this reduces to the integral which can be calculated using the saddle point approximation
\beq
\label{pipoint4}
\Expect{\R^4}_{\mu_0\dot{\pi}^4}&=&\left[-{M_4^4\over \mpl^2 \dot{H}}\right]{6\mu_0\alpha^{9/2}\sqrt{2\pi}c_s^2 \Delta_\zeta^3} \frac{1}{k_1k_2k_3k_4K^5}\sin\left(\alpha\ln K/k_\star\right)\ .
\eeq
The explicit calculation confirms the qualitative conclusions of section \ref{simplest}: the four point function scales as $\alpha^{9/2}\Delta_\zeta^3$. In the folded limit there is an enhancement of oder $\alpha$. 

Let us consider the ratio of the signal-to-noise with respect to the 2-point function. If we ignore the contribution from the folded limit, we obtain
\be
\frac{\langle\zeta^4\rangle/\langle\zeta^2\rangle^2}{\delta\langle\zeta^2\rangle/\langle\zeta^2\rangle}\sim \frac{M_4^4}{\mpl^2\dot H}c_s^2\alpha^4\zeta^2\sim \frac{\omega^4}{\Lambda_{U,4}^4} \lesssim  1\ .
\ee
Therefore the signal to noise ratio of the four point function will be small. The folded limit can give a further enhancement  that is  proportional to  $\alpha^{1/2}$ as for the 3-point function discussed in the previous section. We expect the numerical coefficient for this scaling, that was very small for the 3-point function, to be even smaller in the case of the 4-point function because we expect that the fraction of parameter space affected by the folded limit to be numerically smaller in the case of the 4-point function. Because  $\alpha^{1/2}$ is a slowly growing function we expect the value of signal to noise to become large (compared with the one for the 2-point function) only in the marginal area of $\alpha\sim \alpha_{\rm saturation}$ where the effective theory is becoming strongly coupled. 

We finally add a comment on the multifield inflationary case. As discussed in~\cite{Senatore:2010wk}, in the case of multifield inflation it is possible to have a quartic operator of the form $(\partial_i\sigma)^4$, with $\sigma$ being an additional scalar field, without the presence at the same time of any other cubic or quartic terms. This is impossible in the context of single clock inflation~\cite{Senatore:2010jy}. If we suppose that the coefficient of this operator or of the quadratic Lagrangian  have an oscillatory component, then we may ask ourselves if this quartic operator can generate large oscillating non-Gaussianities. The answer is again no. Indeed, under renormalization, this operator would induce an oscillating component in the part of the quadratic Lagrangian proportional to $(\partial_i\sigma)^2$, without renormalizing the time kinetic part. The resulting linear equation of motion would be different than the one we studied in sec.~\ref{mode_function}, with the oscillating term simply sitting in front of the spatial kinetic term. Apart for this detailed difference, the effect of the 2-point function is of the same order as the one found before, as it can be verified by explicitly finding the solution for the wavefunction. So, we conclude that also in this case the leading signal is on the 2-point function. A full exploration of the multifield inflationary scenario and of all its symmetries lies beyond the scope of the present paper.

\section{Conclusions}

Non-Gaussianities provide one of the most important probes of Physics of the Early Universe. They are closely connected to the interacting structure of the inflationary Lagrangian and their signal is not strongly constrained by the symmetries of the problem, as it happens for the 2-point function. This leaves room for a large amount of non-trivial information to be encoded in higher order correlation functions. For these reasons their detection would represent a  huge improvement in our capabilities to probe inflation. It is therefore of the utmost importance to understand  all possible non-Gaussian signatures that can be generated during Inflation. In this context the Effective Field Theory (EFT) of Inflation represents the ideal setup, as it is particularly well-suited for exploring the whole parameter space of Inflationary models, and classifying them in terms of symmetries of the underlying Lagrangian. The EFT of Inflation also naturally provides an insight into the scaling and the relative importance of the various operators. Different theories can indeed be classified according to the different symmetries they respect, very similarly to what happens in the particle physics context.

In this paper we have considered the case in which the Goldstone boson of time translations is endowed with a softly broken discrete shift symmetry, and we have studied the observational consequences of this pattern of symmetry breaking. Our study unifies and generalizes previously known inflationary models such as Axion Monodromy Inflation.  Because of the non-linear realization of time-diff.s, the discrete shift symmetry forces the presence of parameters in the EFT Lagrangian that oscillate with time. This induces a resonance as the modes expand inside the horizon. The resulting leading signature is the presence of oscillatory features in the power spectrum of density fluctuations. 

Additionally, the same symmetries force the presence of interacting operators for the Goldstone boson that, when combined with the resonant effects, lead to non-Gaussianities with remarkable oscillatory features. Within the regime of validity of the effective theory and for technically natural theories, we find that the signal to noise ratio of this non-Gaussian features (which represents how large these non-Gaussianities are) is always smaller than the one in the 2-point function. That is to say that analysis in search of these oscillatory features can be safely restricted to the 2-point function. Higher-order correlation functions induced by oscillatory operators can be detected only if the oscillatory 2-point function is detected with high signal to noise, or if some (unexpected) degeneracies are present in the 2-point function. Such a detection would be a remarkable confirmation that Inflation actually took place, and that the Goldstone boson was endowed with a discrete shift symmetry.

Our bound originates from the fact that within the Effective Theory there is a maximum energy scale $\Lambda_U$ beyond which unitarity is violated. Non-Gaussianities are small because resonance frequencies larger than $\Lambda_U$ cannot be considered.  It would be interesting to see if our conclusions remain unaltered when one considers specific UV completions of the Effective Theory, where, probably after introducing new degrees of freedom, one could explore regimes of energies larger than $\Lambda_U$. As we briefly argue in the main part of the paper, the possibility that our conclusions might change seems unlikely to us. A definitive answer can however come only from studies of detailed UV completions.

\section*{Acknowledgement}

We thank D. Baumann, S. Dubovsky, R.~Flauger, D.~Green, A.~Gruzinov, J.~Kaplan, J.~Maldacena, L.~McAllister, E.~Pajer, S.~Shenker, E.~Silverstein, J. Wacker, and M. Zaldarriaga for useful discussions. A.D. thanks the Stanford Institute for Theoretical Physics for hospitality.
 The research of S.R.B  was supported by the US DOE under contract number DE-AC02-76SF00515. The research of A.D. was supported by the DOE grant DE-FG02-90ER40542, by Monell Foundation, and in part by the grant RFBR 07-02-00878 and the Grant for Support of Scientific Schools NSh-3035.2008.2. M.M. was supported by the Mark Leslie graduate award.

%%%%%%%%%%%%%%%%%%%%%%%%%%%%%%%%%%%%%%%%%%%%%%%%%%%%%%%%%%%%%%%%%%
%%%%%%%%%%%%%%%%%%%%%%%%%%%%%%%%%%%%%%%%%%%%%%%%%%%%%%%%%%%%%%%%%%
%%%%%%%%%%%%%%%%%%%%%%%%%%%%%%%%%%%%%%%%%%%%%%%%%%%%%%%%%%%%%%%%%%

\appendix

\section{Mixing with Gravity\label{app:mixing}}

In all the calculations we have always concentrated on the decoupling limit and we neglected the mixing with gravity. This was due to the intuitive fact that the resonance frequency was way inside the horizon when non-Gaussianities are large: $\alpha\gg 1$, and so we expect that mixing with gravity should be suppressed either by the ratio of these two scales, which is indeed $1/\alpha$, or by the slow roll coefficients. In order to check our intuition, we write the metric using the ADM parametrization
\be
ds^2=-N^2 dt^2+\hat g_{ij}\left(dx^i+N^i dt\right)\left(dx^j+N^j dt\right)\ ,
\ee
we neglect tensor modes, and we choose the spatially flat gauge
\be
\hat g_{ij}=a^2\delta_{ij}\ .
\ee
For simplicity we ignore tensor fluctuations because they will be negligible. Now we should solve the constraint ADM parameters $N,\, N^i$ at all orders in $\pi$~\footnote{Recall that to obtain the action for $\pi$ at the $n^{th}$ order one just needs to solve for the $N,N_i$ up to order~$(n-2)$\cite{Maldacena:2002vr}.}, and plug them into the first line of the action (\ref{eq:actiontad}). In this gauge the action reads
\bea\label{eq:action-contraint}
&&S=\int d^4x\;a^3\mpl^2\left\{\frac{1}{2N}(E_{ij}^2-E^2)\right.\\ \nonumber
&&\qquad+\left.\Hdot\left[-N^{-1}(1+\dot\pi)^2+2N^{-1}(1+\dot\pi)N^i\partial_{i}\pi+N\left(\partial_{i}\pi\right)^2-N^{-1}(N^i\partial_{i}\pi)^2\right]\right.\\ \nonumber
&&\qquad\left.-N(3\Htwo+\Hdot)\right\}\,,
\eea
where
\beq
E_{ij}=\frac{1}{2}\dot{\hat g}_{ij} - N_{(i;j)}=a^2 H\delta_{ij}- N_{(i,j)}\ ,\quad N_{(i;j)}={1\over2}(N_{i;j}+N_{j;i})\,,
\eeq
and $;$ stands for the covariant derivative with respect to the spatial metric $\hat g_{ij}$, which in this gauge is simply an ordinary derivative.  In this gauge, the constraint equations read
\bea
\label{constraint_i}
&&\partial_j[N^{-1}(E^j_i-\delta^j_i E)]+2N^{-1}\Hdot[(1+\dot\pi)\partial_{i}\pi-N^j\partial_{j}\pi\partial_{i}\pi]=0\,,\\
\label{constraint}
&&R^{(3)}-\frac{1}{2N^2}(E_{ij}^2-E^2)-(3\Htwo+\Hdot)\quad\nonumber\\
&&+\Hdot[N^{-2}(1+\dot\pi)^2-2N^{-2}(1+\dot\pi)N^i\partial_{i}\pi+\left(\partial_{i}\pi\right)^2+N^{-2}(N^i\partial_{i}\pi)^2]=0\,.
\eea
We are interested in solving these equations perturbatively in $\pi$. It is therefore useful to re-arrange the equations in a perturbative fashion as follows
\beq
\label{N}
\delta N_{,i}&=&-N\frac{\Hdot}{H(t)}[(1+\dot\pi)\partial_{i}\pi-N^j\partial_{j}\pi\partial_{i}\pi]\nonumber\\
&&\quad-\frac{N^2}{2H(t)}\partial_{j}[N^{-1}(\delta^j_iN^k_{,k}-\frac{1}{2}(N^j_{,i}+N^{,j}_i)]\,,\\
\label{Ni}
N^i_{,i}&=&\frac{\Hdot}{2H(t)}[(1+\dot\pi)^2-N^2-2(1+\dot\pi)N^i\partial_{i}\pi+N^2(\partial_{i}\pi)^2+(N^i\partial_{i}\pi)^2]\nonumber\\
&&\quad +\frac{3}{2H(t)}\left(H(t)^2-N^2\Htwo]\right)-\frac{1}{4H(t)}(N_{(i,j)}^2-(N^i_{,i})^2)\,,
\eeq
where we have defined $N=1+\delta N$. Note that only the terms that come from the the time-dependent coefficients of tadpoles $\sqrt{-g}\delta g^{00}$ and $\sqrt{-g}$ have an explicit $t+\pi$ dependence as the Einstein-Hilbert action is invariant on its own. The first order solution is
\beq\label{eq:Nisol}
\delta N^{(1)}= -\frac{\dot H}{H}\pi\,,\qquad N^{i\;(1)}{}_{,i}=\frac{\dot H}{H^2}\partial_t(H\dot\pi)\,.
\eeq
It is not easy to solve equations \eqref{N} and \eqref{Ni} beyond linear order. However it is straightforward to understand the order in $\alpha$ at which they contribute. From the structure of the equations we see that at each new order we get only one extra derivative. This extra derivative either comes from explicit derivatives of $\pi$, or from Taylor expansion of $\Hdot$ and $\Htwo$, or substitution of lower order solutions for $N$ and $N^i$ in the r.h.s. The leading $n^{th}$~order solution therefore scales as
\beq
\delta N^{(n)}\sim N^{i\; (n)} \sim \ep \epsosc \alpha^{n-1}H^n\pi^n\,.
\eeq
%\footnotetext{{\bf ?!?!?!?!?!?!?!? Remove?!?!?!?!?!?!}To analyze the $\pi$-$\zeta$ conversion it is useful to know the super-horizon behavior of $N$ and $N^i$ where spatial derivatives go to zero exponentially, and time derivatives of $\pi$ are slow-roll suppressed. It follows from \eqref{N} and \eqref{Ni} that in that limit
%\beq
%\label{N_super}
%N^{(n)}\sim \frac{H^{(n)}}{H}\pi^n\,,\\
%\label{Ni_super}
%N^i\sim \frac{H^{(n)}}{H}\frac{\partial_i}{\d^2}\pi^{n-1}\dot\pi\,.
%\eeq
%}

Interaction terms due to mixing with gravity will originate from substituting the solutions for $N,N^i$ into the action (\ref{eq:action-contraint}). At a given order in $\pi$, we can compare the interactions due to mixing with gravity with the leading one of order $\mpl^2 H^{(n+1)} \pi^n$. Going through all  interactions is quite tedious, and we simply consider the two leading terms. One comes from the interaction operator associated with
\be
{\cal L}_{(\partial_i N_j)^2}\sim \mpl^2(\partial_j N_i)^2\sim \mpl^2 \frac{H^{(m)}H^{(n-m)}}{H^2} \pi^{n-2}(\partial_i\pi)^2\ ,
\ee
which is suppressed with respect to the leading one
\be
{\cal L}_{H^{(n+1)}\pi^n}\sim \mpl^2 H^{(n+1)}\pi^n\ ,
\ee
by the factor
\be
\frac{{\cal L}_{(\partial_i N_j)^2}}{{\cal L}_{H^{(n+1)}\pi^n}}\sim \epsilon\,\epsosc\,\lesssim \epsilon^2\ .
\ee

Another interaction that scales parametrically in a different  way comes from terms of the form
\be
{\cal L}_{\dot H(t+\pi)\delta N }\sim \epsilon H H^{(n)}\pi^n\ ,
\ee
where we have taken the linear term in $\delta N$ as the non-linear terms in $\delta N$ would give rise to further suppressed interactions. This compares to the leading term as
\be
\frac{{\cal L}_{\dot H(t+\pi)\delta N }}{{\cal L}_{H^{(n+1)}\pi^n}}\sim \frac{\epsilon}{\alpha}\ .
\ee
Finally, it is possible to check that diagrams where the graviton is exchanged, are even more suppressed than the one we consider.
We therefore conclude that mixing with gravity is suppressed by ${\rm Min}(\epsilon/\alpha,\epsilon\,\epsosc)\ll \epsilon$.

%%%%%%%%%%%%%%%%%%%%%%%%%%%%%%%%%%%%%%%%%%%%%%%%%%%%
%%%%%%%%%%%%%%%%%%%%%%%%%%%%%%%%%%%%%%%%%%%%%%%%%%%%
%%%%%%%%%%%%%%%%%%%%%%%%%%%%%%%%%%%%%%%%%%%%%%%%%%%%

\section{$\pi$ to $\R$ Conversion and Field Redefinition\label{app:redefinition}}

In the EFT of Inflation we computed $n$-point functions of the Goldstone boson $\pi$, which is the field that naturally manifests the decoupling limit. 
However, we are interested in correlation functions of the curvature perturbation $\zeta$, which has the useful property of being time-independent for modes outside the horizon.  In order to convert $\pi$ $n$-point functions into $\zeta$ $n$-point functions, we need to find the non-linear relations between $\pi$ and $\zeta$ at the required level. In this Appendix we show that only the linear relationship $\zeta=-H\pi$ is important for $n$-point functions with detectable signal to noise ratio.

%\section*{\large Higher order gauge transformation}
 %\partial_{i} T\partial_{i} T(-N^2+\e^{2(\rho(t+T))}N^{m}N_{m})%%%%%%%%%%%%%%%%%%%%%%%%%%%%%%%%%%%%%%%%%%%%%%%%%%
%%%%%%%%%%%%%%%%%%%%%%%%%%%%%%%%%%%%%%%%%%%%%%%%%%%%
%%%%%%%%%%%%%%%%%%%%%%%%%%%%%%%%%%%%%%%%%%%%%%%%%%%%

\subsection{The non-linear relation}
In order to obtain the non-linear relation between $\pi$ and $\zeta$, we need to perform a gauge transformation from any gauge where $\pi$ is defined to the unitary gauge where $\zeta$ is defined and where $\pi$ is set to zero. Notice that apart for terms coming from the mixing with gravity (which are subleading), we could choose to describe $\pi$ in any gauge distant enough from the unitary gauge~\cite{Cheung:2007st,Baumann:2010tm}. Explicit results will differ by subleading terms coming from the mixing with gravity. We can choose for example  to define $\pi$ in the spatially flat gauge:
\beq
g_{ij}=a(t)^2 \delta_{ij}\,.
\eeq
Here and throughout we neglect tensor modes, as they do not change the answer at leading order. 
In the unitary gauge $\pi=0$ and the scalar degree of freedom is encoded in the metric:
\beq
\label{unit}
g_{ij}=a(t)^2 e^{2\zeta}\delta_{ij}\,.
\eeq
In order to set $\pi=0$, we perform a time diffeomorphism
\beq
\label{diff}
\tilde t = t+T(t,x)\,,
\eeq
where $\tilde t$ is the time variable of the $\pi$-gauge, $t$ is that of the unitary gauge and $T(t,x)$ can be found iteratively from
\beq
\label{T}
\pi(\tilde t,x)+T(t,x)=0\,.
\eeq
For example this gives at second order
\beq
T = -\pi +\pi\dot\pi + {\cal O}(\pi^3)\,.
\eeq
In the ADM parametrization, the metric in the spatially-flat gauge is of the form
\beq
\label{ds}
d\tilde s ^2 = -N^2 d\tilde t^2+ a^{2}(\tilde t)\delta_{ij}\left(dx^i+N^id\tilde t\right)\left(dx^j+N^jd\tilde t\right)\,,
\eeq
where $N$ and $N^i$ are given by equations (\ref{N},\ref{Ni}). Applying the diffeomorphism \eqref{diff} to~\eqref{ds}, the spatial part of the metric becomes
\beq\label{gij}
g_{ij}=a^2(t+T)\left[\delta_{ij}+\partial^{}_{(i}T\delta^{m}_{j)}N_m+\partial_{i} T\partial_{j} T\left(-\frac{1}{a^2(t+T)}N^2+N^{m}N_{m}\right)\right]\,,
\eeq
where $N_i=N^i$. This metric is not of the form of \eqref{unit}, due to some second and higher order terms in $\pi$. Therefore one is forced to make an additional spatial diffeomorphism $\tilde{x}^i=x^i+\epsilon^{i}(t,x)$.

It is very difficult to derive explicitly the full non-linear spatial diff. necessary to reach the $\zeta$-gauge. However, since we need to convert $\pi$ fluctuations to $\zeta$ fluctuations only  when all modes are well outside the horizon, we can restrict ourselves only to leading order in the spatial derivatives. This simplifies greatly the derivation because, as we can see, the last extra terms in (\ref{gij}) vanish in the limit that the spatial derivatives for all fields vanish~\footnote{While the terms in $T$ do not contain any spatial derivative, this is not the case for $N_i$, which must start at least at first order in spatial derivatives. We are going to prove by induction in the order of the fluctuations this quite intuitive fact. We can imagine to perform the diff.  $\tilde{x}^i=x^i+\epsilon^{i}(t,x)$ and expand $\epsilon_i$ perturbatively in fluctuations: $\epsilon_i^{\{n\}}$ being of order $n$ in the fluctuations. Notice that $\epsilon_i$ starts at second order. Then we can imagine  to solve iteratively in $\epsilon_i^{\{n\}}$. Indeed, similarly to what is done in~\cite{Maldacena:2002vr}, the condition that $\epsilon_i^{\{n\}}$ needs to satisfy is 
\be\label{eq:spatial}
\delta h^{\{n\}}_{ij}+\partial_i\epsilon^{\{n\}}{}^j+\partial_j\epsilon^{\{n\}}{}^i+\ldots=\beta^{\{n\}}\delta_{ij} \ ,
\ee
where $\delta h^{\{n\}}_{ij}$ represents the $n$-th order part of the last three terms in (\ref{gij}) and where $\ldots$ stands for terms involving lower order terms in $\epsilon_i^{\{n\}}$: $\epsilon_i^{\{n-1\}}\,,\,\epsilon_i^{\{n-2\}}\,\ldots$ or analogously in $\beta$. By taking the trace and by applying $\partial_i\partial_j$ to (\ref{eq:spatial}), we can solve for $\beta^{\{n\}}$ to be schematically
\be
\beta^{\{n\}}\sim\delta h^{\{n\}}_{ii}+\frac{\partial_i\partial_j}{\partial^2}\delta h_{ij}^{\{n\}}+\ldots\ ,
\ee
where $\ldots$ contains lower order terms and where we have neglected numerical factors since it will turn out that we are  not interested in the actual explicit solution. Expanding in smallness of the spatial derivatives, if we assume that the lower order terms have the property that $N^i$ and $\epsilon_i$ are at least of first order in the spatial derivatives and $\beta$ is at least of second order, then we see that  the $n^{th}$ order $\beta^{\{n\}}$ is of second order in spatial derivatives as well, and similarly upon substitution in (\ref{eq:spatial}) we conclude that also $\epsilon_i^{\{n\}}$ is of first order in spatial derivatives. With this information we are now going to show that also $N_i^{(n)}$ start at first oder in the derivatives. Indeed in~\cite{Maldacena:2002vr} it is shown to all orders that $N_i^{(\zeta)}$ in $\zeta$ gauge starts at first oder in the spatial derivatives, and upon performing the spatial diff. parametrized by $\epsilon_i$, it is easy to see that at this point also $N_i$ in $\pi$-gauge starts at first order. From this we see that assuming that at the orders $1,\ldots,n-1$, $N_i$ and $\epsilon_i$ are of first order and $\beta$ is of second order in spatial derivatives, we obtain that the same condition holds for the $n^{th}$ order. Since in (\ref{eq:Nisol}) we show explicitly that this counting of derivatives holds at the lowest orders in the fluctuations, we can conclude that this holds for all the orders, as we wanted to show. Explicit inclusion of spatial diff.s would modify the relations  between $\zeta$ and $\pi$ to 
\be
\zeta^{\{n\}}=\left.\ln \frac{a(t+T)}{a(t)}\right|^{\{n\}}+\beta^{\{n\}}+\ldots
\ee
with $\ldots$ representing terms containing lower order terms in $\beta$ and $T$. Eventiually we conclude that spatial diff.s can be safely neglected in the redefinition between $\zeta$ and $\pi$.
}.
This implies that in this limit the spatial diff. is irrelevant, and the relation between $\pi$ and $\zeta$ reduces to
\beq
\label{convert}
\zeta=\ln \left(\frac{a(t+T)}{a(t)}\right)\ ,
\eeq
which can be combined with (\ref{T}) to obtain, for example up to quadratic order
\beq
\label{zeta_pi_2}
\zeta=-H\pi+\frac{1}{2}\dot H \pi^2 + H\pi\dot\pi+{\cal O}(\pi^3)\ .
\eeq
Notice that due to the mixing with gravity $\pi$ is time dependent even after horizon crossing. By using the relation $\zeta=-H\pi$ and the fact that $\zeta$ is constant outside of the horizon, we derive $\dot\pi\sim \epsilon H\pi$. This tells us that the time dependence can be neglected at leading order in the slow roll parameters.
We conclude that for our purposes the effect of the non-linear relation between $\zeta$ and $\pi$ has the following scaling 
\beq
\label{zeta_pi}
\zeta^{\{n\}} \sim H^{(n-1)}\pi^n\ \sim \epsilon\,\epsosc \alpha^{n-2}\zeta^n\,\ ,
\eeq
where $\zeta^{\{n\}} $ stays for the contribution of order $n$ in powers of $\zeta$.

\subsection{Corrections due to non-linear $\pi$ to $\zeta$ conversion}

We are now going to show that the effect of the field redefinition arising from the oscillatory part beyond linear order is irrelevant for $n$-point functions that have a non-negligible signal to noise. In order to do this, we can use the constraint $\epsosc\lesssim \epsilon \alpha^{1/2}$ from (\ref{eq:epsoscv}) to bound (\ref{zeta_pi}) to be
\be
\zeta^{\{n\}}\lesssim \epsilon^2\alpha^{n-5/2}\zeta^n\ .
\ee 
There is also the field redefinition of $\pi$ in eq.~(\ref{redefine}) to take into account. In terms of $\zeta$ correlation functions this is equivalent to the field redefinition
\be
\zeta^{\{n\}}\sim \frac{H^{(n)}}{H^{n-1}\dot H}\zeta^n\sim\epsosc\alpha^{n-1}\zeta^n\lesssim \epsilon\alpha^{n-3/2}\zeta^n\ .
\ee
We see that the second field redefinition is more important than the first by a factor $\alpha/\epsilon\gg 1$, and so we concentrate on that.

Let us see how the redefinition contributes to an $n$-point correlation function that we can denote by~$\langle\zeta^{n}\rangle_{red.}$:
\be\label{eq:tempestimate}
\langle\zeta^{n}\rangle_{red}=\sum_{m_1}\dots \sum_{m_n}\langle\zeta^{\{m_1\}}\dots\zeta^{\{m_n\}}\rangle\ .
\ee
Since for the consistency of the effective theory we have $\alpha\ll \zeta^{-1/2}$, we see that the leading effect comes from substituting the lowest order possible field redefinition. Further, the structure of the field redefinition with $\epsilon\ll1$ and $\alpha\gg1$ suggests that the leading term is obtained by replacing only one of the $\zeta$'s with the quadratic field redefinition. We therefore have
\be\label{eq:tempestimate2}
\langle\zeta^{n}\rangle_{red}\sim \epsosc\alpha \langle\zeta^{n+1}\rangle\ .
\ee
If $n$ is odd, this is the leading effect; if $n$ is even, we have to either go to higher order in the field  redefinition or to compute the non-Gaussian correlation. Let us start with $n$ odd.  The signal to noise associated to this field redefinition is given by
\be
\frac{\langle\zeta^{n}\rangle_{red}}{\Rtwo^{n/2}}\sim \epsosc \alpha\Rtwo^{1/2}\lesssim \epsilon \alpha^{1/2} \Rtwo^{1/2}\ ,
\ee
where we used that $\epsosc\lesssim \epsilon \alpha^{-1/2}$. This is still less than $\zeta$ taking the relation $\alpha\ll \zeta^{-1/2}$ into account.
 In order to be detectable in the foreseeable future, we need this ratio to be larger than $10^{-5}$ (corresponding to the optimistic case of signal to noise capable of detecting $f_{NL}\sim 1$). By a coincidence this is of order  $\Rtwo^{1/2}$. Therefore the effect from the field redefinition is too small.

If $n$ is even, one way is to insert either a third order field redefinition, leading to a signal to noise of order
\be
\frac{\langle\zeta^{n}\rangle_{red}}{\Rtwo^{n/2}}\sim \epsosc \alpha^2 \Rtwo^{1/2}\lesssim \epsilon\alpha^{3/2} \Rtwo\lesssim \epsilon \Rtwo^{5/4}\ ,
\ee
where in the second inequality we used the bound on $\alpha \ll \zeta^{-1/2}$ coming from the consistency of the effective theory. This is again too small to be detectable. Alternatively we can use a non-Gaussian correlation function. From (\ref{R_N}), this is schematically of the form
\beq
\label{R_n_scaling}
\Expect{\R^n}\sim \epsosc \alpha^{2n-7/2}\Rtwo^{n-1}\, ,
\eeq
which leads to a signal to noise of order
\be
\frac{\langle\zeta^{n}\rangle_{red}}{\Rtwo^{n/2}}\sim \epsosc^2 \alpha^{2n-1/2} \Rtwo^{n/2}\lesssim \epsilon^2 \Rtwo^{3/8}\ , 
\ee
which is again too small to be detectable.
We conclude that the effect of the field redefinition is too small to lead to a detectable signal.

The effect of the multivertex diagrams and the collective effect of the lower order terms neglected above were discussed in \cite{Pajer}. However if the ratio of the resonance frequency with respect to the cutoff is as large as half, which is somewhat border line, then we can have up to the $10^{th}$-point function possible detectable, which is about the same value at which multi-vertex diagrams tend to become important~\cite{Pajer}.

%%%%%%%%%%%%%%%%%%%%%%%%%%%%%%%%%%%%%%%%%%%%%%%%%%%%%%%%%%%%%%%%
%%%%%%%%%%%%%%%%%%%%%%%%%%%%%%%%%%%%%%%%%%%%%%%%%%%%%%%%%%%%%%%%
%%%%%%%%%%%%%%%%%%%%%%%%%%%%%%%%%%%%%%%%%%%%%%%%%%%%%%%%%%%%%%

\section{\label{signal}Signal-to-Noise ratio for 3-point function}

To estimate the signal to noise ratio of the 3-point function we need to calculate 
\beq
\label{SN3}
&&\left(\frac{S}{N}(\Expect{\R^3})\right)^2 \simeq V^N\int \frac{d^3\mbf{k}_1}{(2\pi)^3} \frac{d^3\mbf{k}_2}{(2\pi)^3}\frac{d^3\mbf{k}_3}{(2\pi)^3}\frac{\Expect{\R_{\mbf{k}_1}\R_{\mbf{k}_2}\R_{\mbf{k}_3}}\Expect{\R_{\mbf{k}_1}\R_{\mbf{k}_2}\R_{\mbf{k}_3}}}{\Expect{\R_{\mbf{k}_1}\R_{\mbf{k}_2}\R_{\mbf{k}_3}\R_{\mbf{k}_1}\R_{\mbf{k}_2}\R_{\mbf{k}_3}}}   \nonumber\\
&&=V^N\int \frac{d^3\mbf{k}_1}{(2\pi)^3}
\frac{d^3\mbf{k}_2}{(2\pi)^3} 
\frac{d^3\mbf{k}_{3}}{(2\pi)^3}
(2\pi)^3\delta^3(\mbf{k}_1+\mbf{k}_2+\mbf{k}_3)
\frac {A_3^2 B_3^2(k_i)(2\pi)^3\delta^3(\mbf{0})}
{((2\pi)^3\delta^3(\mbf{0})\Rtwo)^3\prod_ik_i^{-3}}\ ,
\eeq
where $A_n$ and $B_n$ are given in \eqref{R_N} and where for simplicity we neglect the transfer functions and sky projection effects because we expect them not to affect the result significantly. Taking $(2\pi)^3\delta^3(\mbf{0})=V$ we get
\beq
\label{SN31}
\left(\frac{S}{N}\left(\Expect{\R^n}\right)\right)^2\simeq \frac{A_n^2}{\Rtwo^n} V \int \frac{d^3\mbf{k}_1}{(2\pi)^3}\frac{d^3\mbf{k}_2}{(2\pi)^3} \frac{d^3\mbf{k}_{3}}{(2\pi)^3}(2\pi)^3\delta^3(\mbf{k}_1+\mbf{k}_2+\mbf{k}_3)\frac{B_3^2}{\prod_ik_i^{-3}}\,.
\eeq
Since the integrand depends only on the magnitude of $k_1,k_2,k_3$ we can use the delta function to eliminate angular variables of the last two integrals 
\beq
\int \frac{d^3\mbf{k}_2}{(2\pi)^3} \frac{d^3\mbf{k}_{3}}{(2\pi)^3}(2\pi)^3\delta^3(\mbf{k}_1+\mbf{k}_2+\mbf{k}_3)=\frac{3!}{4\pi^2 k_1}\int_{k_1/2}^{k_1} k_2dk_2 \int_{k_1-k_2}^{k_2}k_3 dk_3\,.
\eeq
Here we have also used the symmetry of the integrand with respect to interchange of $k_1,k_2,k_3$, multiplied by $3!$, and limited the integrals to the region $k_3<k_2<k_1$.

Using $x_{2,3}\equiv k_{2,3}/k_1$, $B_3(k_i)$ at leading order in $\alpha$ is given by 
\beq
B_3(k_i)&=& \frac{1}{ \prod_i k_i^2} \left[ \sin\left(\alpha \ln(K/k_\star)\right)+O\left(\frac1\alpha\right)\right]\nonumber\\
&\equiv& \frac{1}{k_1^6}b_3(x_2,x_3,k_1/k_*)\simeq\frac{1}{k_1^6} \frac{\sin\left(\alpha \ln\left((1+x_2+x_3)k_1/k_\star\right)\right)}{x_2^2x_3^2}\,,
\eeq
where $b_3$ is a scale invariant function except for the sine term. Changing the variables of integration from $k_2,k_3$ to $x_2,x_3$ in \eqref{SN31} we obtain
\beq
\label{S_N32}
\left(\frac{S}{N}\left(\Expect{\R^3}\right)\right)^2&\simeq& \frac{3A_3^2}{2\pi^2\Rtwo^3} V \int \frac{d^3\mbf{k}_1}{(2\pi)^3} \int_{1/2}^{1} dx_2 \int_{1-x_2}^{x_2}dx_3 x_2^4 x_3^4 b_3^2(x_2,x_3,k_1/k_*)\nonumber\\
&=&\frac{3A_3^2}{2\pi^2\Rtwo^3} V \int \frac{d^3\mbf{k}_1}{(2\pi)^3} I(k_1/k_*)\,,
\eeq
where $I(k_1/k_*)$ is the result of the integrals over $x_2$ and $x_3$ and in principle depends on $k_1$. However, since the dependence on $k_1$ solely comes from  $\sin^2(\alpha \ln\left(\left(1+x_2+x_3\right)k_1/k_*\right))$ in $b^2_3$, for large $\alpha$'s it has already been approximately averaged to 1/2 by the $x_2$ and $x_3$ integrals. Therefore $I(k_1/k_*)$  can be approximated to be a $k_1$ independent constant
\beq
I(k_1/k_*)\simeq  I(1)\simeq \frac{1}{2}\int_{1/2}^{1} dx_2 \int_{1-x_2}^{x_2}dx_3=1/8\, .
\eeq

Similarly we can calculate the signal to noise ratio for the modifications of the 2-point function. Using
\beq
B_2(k)\simeq \frac{2}{k^3}\sin(\alpha\ln(2k/k_*))\,,
\eeq
we have
\beq
\left(\frac{S}{N}\left(\delta\Expect{\R^2}\right)\right)^2&\simeq& \frac{A_2^2}{\Rtwo^2} V \int \frac{d^3\mbf{k}_1}{(2\pi)^3} 4\sin^2(\alpha\ln(2k/k_*))\nonumber\\
&\simeq& \frac{2 A_2^2}{\Rtwo^2} V \int \frac{d^3\mbf{k}_1}{(2\pi)^3}\,,
\eeq
where in the second line we have replaced $\sin^2(\alpha\ln(2k/k_*))$ by its average. 

Therefore the signal to noise ratio in 3-point function compared to the modifications of the 2-point function becomes 
\beq\label{eq:signallast}
\frac{\left(\frac{S}{N}\left(\Expect{\R^3}\right)\right)}{\left(\frac{S}{N}\left(\delta\Expect{\R^2}\right)\right)} = \frac{\sqrt{3}I^{1/2}(1)A_3}{2\pi \Rtwo^{1/2}|A_2|} \simeq 
 \frac{\sqrt{3}\alpha^2\Rtwo^{1/2}}{4\sqrt{2}\pi}\,.
\eeq
Even by taking $\alpha$ so high it reaches its upper bound \eqref{eq:alphaboundprec} the ratio (\ref{eq:signallast}) reaches its maximum $\sqrt{3}/2\sim 0.87$. For smaller $\alpha$'s, this number scales as $\alpha^2/\alpha_{\rm saturation}^2$, and as long as the EFT is not strongly coupled it is expected to be much smaller than that.

Similar techniques can be used to evaluate the signal to noise ratio for the 3-point functions of section \ref{3pt}. The results are plotted in Fig.~\ref{Fig: S_N}.

%%%%%%%%%%%%%%%%%%%%%%%%%%%%%%%%%%%%%%%%%%%%%%%%%%%%%%%%%%%%%%%%
\section{Oscillating models and standard shapes}\label{app:cosines}

%%%%%%%%%%%%%%%%%%%%%%%%%%%%%%%%%%%%%%%%%%%%%%%%%%%%%%%%%%%%%
\begin{figure}[htbp]
\begin{center}
\includegraphics[width=2.8in]{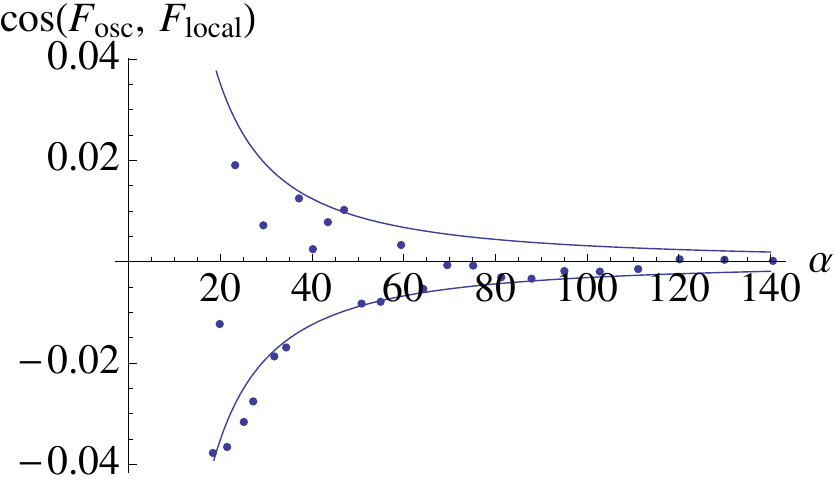}
\includegraphics[width=2.8in]{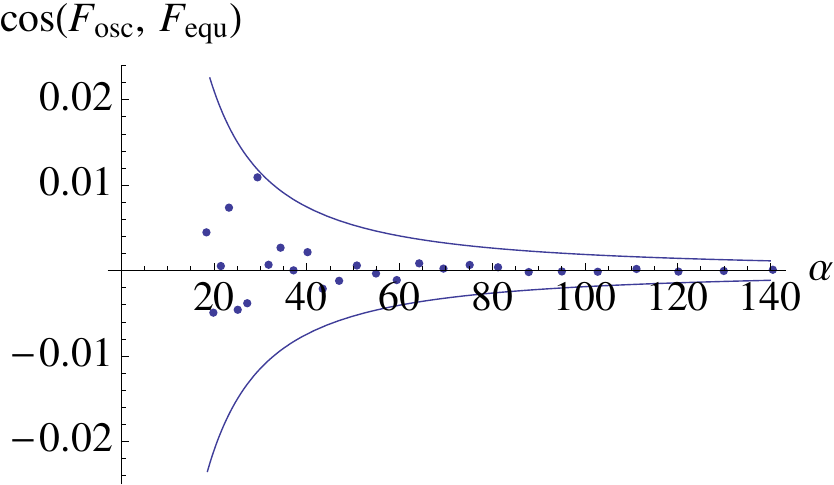}
\includegraphics[width=2.8in]{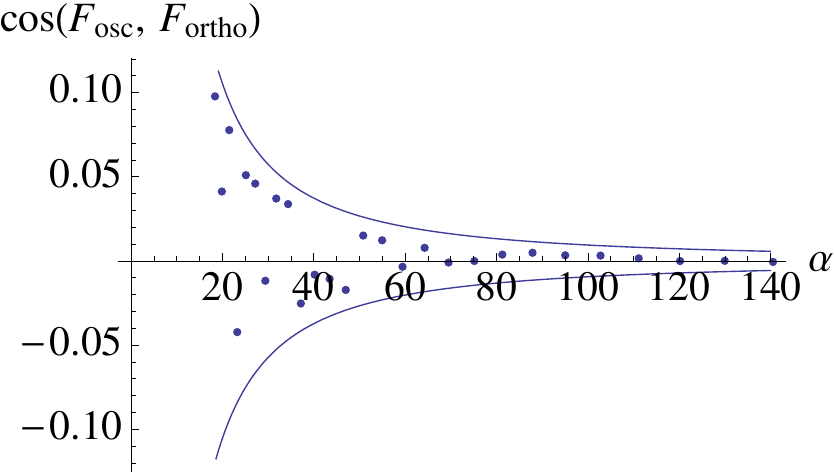}
\caption{\label{Fig:Cosin} Clockwise from top left we have plotted the cosine with local, equilateral and orthogonal shapes, in the valid range of $\alpha$. We see that as $\alpha$ becomes large, the cosine goes to zero.}
\end{center}
\end{figure}
%%%%%%%%%%%%%%%%%%%%%%%%%%%%%%%%%%%%%%%%%%%%%%%%%%%%%%%%%%%%%%%%

Here we show that rapidly oscillating shapes (large $\alpha$'s) are almost perpendicular to the standard shapes and therefore the standard non-Gaussianty searches can not be used to constrain the model in question~\footnote{We stress that the parametrically leading signal is in the 2-point function, and not in non-Gaussianities, that add very little signal to noise.}. We trivially extend the definition of Cosine in reference \cite{Cosin} to incorporate the fact that oscillating shapes are not exactly scale invariant. The only difference is that we need to integrate over all three momenta in the observable range that here we take to be from $k_{min}=10^{-4}$~Mpc$^{-1}$ to $k_{max}=10^{-1}~$Mpc$^{-1}$.  In  \cite{Flauger2} a different definition for Cosine was used which is better suited to the CMB data analysis, while the definition of~\cite{Cosin} is more similar to the analysis from both the CMB and the Galaxy Survey points of view. In Fig.~\ref{Fig:Cosin}, we  consider the shape
\beq
B_3(k_i)&\equiv& \frac{1}{ \prod_i k_i^2} \left[ \sin(\alpha \ln K/k_\star)+\frac1\alpha \cos(\alpha \ln K/k_\star) \sum_{j,i} \frac{k_i}{k_j} \right]\,,\nonumber
\eeq
and calculate the Cosine with Local, Orthogonal and Equilateral shapes as a function of $\alpha$. Solid lines in the plot are showing the semi-analytic prediction for the cosine as a function of $\alpha$. Indeed using the same analysis as in \cite{Flauger1} it is easy to show that Cosine scales as
\beq
C_{local}\lesssim \frac{\pi}{\alpha^{3/2}}\ .
\eeq

%%%%%%%%%%%%%%%%%%%%%%%%%%%%%%%%%%%%%%%%%%%%%%%%%%%%%%%%%%%%%%%%%%
%%%%%%%%%%%%%%%%%%%%%%%%%%%%%%%%%%%%%%%%%%%%%%%%%%%%%%%%%%%%%%%%%%
%%%%%%%%%%%%%%%%%%%%%%%%%%%%%%%%%%%%%%%%%%%%%%%%%%%%%%%%%%%%%%%%%%


\begin{thebibliography}{10}
\bibitem{Cheung:2007st}
  C.~Cheung, P.~Creminelli, A.~L.~Fitzpatrick, J.~Kaplan, L.~Senatore,
  ``The Effective Field Theory of Inflation,''
  JHEP {\bf 0803}, 014 (2008).
  [arXiv:0709.0293 [hep-th]].

%\cite{Cheung:2007sv}
\bibitem{Cheung:2007sv}
  C.~Cheung, A.~L.~Fitzpatrick, J.~Kaplan and L.~Senatore,
  ``On the consistency relation of the 3-point function in single field
  inflation,''
  JCAP {\bf 0802} (2008) 021
  [arXiv:0709.0295 [hep-th]].
  %%CITATION = JCAPA,0802,021;%%
  
  %\cite{Senatore:2009gt}
\bibitem{Senatore:2009gt}
  L.~Senatore, K.~M.~Smith and M.~Zaldarriaga,
   ``Non-Gaussianities in Single Field Inflation and their Optimal Limits from
  the WMAP 5-year Data,''
  JCAP {\bf 1001}, 028 (2010)
  [arXiv:0905.3746 [astro-ph.CO]].
  %%CITATION = JCAPA,1001,028;%%
%\cite{Peskin:1991sw}


 
%\cite{Senatore:2009cf}
\bibitem{Senatore:2009cf}
  L.~Senatore, M.~Zaldarriaga,
  ``On Loops in Inflation,''
  JHEP {\bf 1012 } (2010)  008.
  [arXiv:0912.2734 [hep-th]].

%\cite{Senatore:2010jy}
\bibitem{Senatore:2010jy}
  L.~Senatore, M.~Zaldarriaga,
  ``A Naturally Large Four-Point Function in Single Field Inflation,''
  JCAP {\bf 1101 } (2011)  003.
  [arXiv:1004.1201 [hep-th]].


%\cite{Creminelli:2006xe}
\bibitem{Creminelli:2006xe}
  P.~Creminelli, M.~A.~Luty, A.~Nicolis and L.~Senatore,
  ``Starting the universe: Stable violation of the null energy condition and
  non-standard cosmologies,''
  JHEP {\bf 0612} (2006) 080
  [arXiv:hep-th/0606090].
  %%CITATION = JHEPA,0612,080;%%

%\cite{Bartolo:2010bj}
\bibitem{Bartolo:2010bj}
  N.~Bartolo, M.~Fasiello, S.~Matarrese, A.~Riotto,
  ``Large non-Gaussianities in the Effective Field Theory Approach to Single-Field Inflation: the Bispectrum,''
  JCAP {\bf 1008 } (2010)  008.
  [arXiv:1004.0893 [astro-ph.CO]].

%\cite{Bartolo:2010di}
\bibitem{Bartolo:2010di}
  N.~Bartolo, M.~Fasiello, S.~Matarrese, A.~Riotto,
  ``Large non-Gaussianities in the Effective Field Theory Approach to Single-Field Inflation: the Trispectrum,''
  JCAP {\bf 1009 } (2010)  035.
  [arXiv:1006.5411 [astro-ph.CO]].

%\cite{Senatore:2010wk}
\bibitem{Senatore:2010wk}
  L.~Senatore, M.~Zaldarriaga,
  ``The Effective Field Theory of Multifield Inflation,''
  [arXiv:1009.2093 [hep-th]].

%\cite{Creminelli:2010qf}
\bibitem{Creminelli:2010qf}
  P.~Creminelli, G.~D'Amico, M.~Musso, J.~Norena, E.~Trincherini,
  ``Galilean symmetry in the effective theory of inflation: new shapes of non-Gaussianity,''
  JCAP {\bf 1102 } (2011)  006.
  [arXiv:1011.3004 [hep-th]].

%\cite{senatore_complletion}
\bibitem{senatore_complletion}
  S.~R.~Behbahani,  M.~Mirbabayi, A.~Gruzinov, L.~Senatore, K.~M.~Smith, M.~Zaldarriaga
  in progress.

%\cite{Baumann:2011dt}
\bibitem{Baumann:2011dt}
  D.~Baumann, L.~Senatore, M.~Zaldarriaga,
  ``Scale-Invariance and the Strong Coupling Problem,''
  JCAP {\bf 1105}, 004 (2011).
  [arXiv:1101.3320 [hep-th]].

%\cite{Baumann:2011su}
\bibitem{Baumann:2011su}
  D.~Baumann, D.~Green,
  ``Equilateral Non-Gaussianity and New Physics on the horizon,''
  JCAP {\bf 1109 } (2011)  014.
  [arXiv:1102.5343 [hep-th]].

%\cite{Baumann:2011nk}
\bibitem{Baumann:2011nk}
  D.~Baumann, D.~Green,
  ``Signatures of Supersymmetry from the Early Universe,''
  [arXiv:1109.0292 [hep-th]]. %\cite{Baumann:2011nm}
%\bibitem{Baumann:2011nm}
  D.~Baumann, D.~Green,
  ``Supergravity for Effective Theories,''
  [arXiv:1109.0293 [hep-th]].


%\cite{Nacir:2011kk}
\bibitem{Nacir:2011kk}
  D.~L.~Nacir, R.~A.~Porto, L.~Senatore, M.~Zaldarriaga,
  ``Dissipative effects in the Effective Field Theory of Inflation,''
  [arXiv:1109.4192 [hep-th]].


\bibitem{McAllister}
  L.~McAllister, E.~Silverstein and A.~Westphal,
  ``Gravity Waves and Linear Inflation from Axion Monodromy,''
  Phys.\ Rev.\  D {\bf 82}, 046003 (2010)
  [arXiv:0808.0706 [hep-th]].
  %%CITATION = PHRVA,D82,046003;%%
  
  
  %\cite{Cornwall:1974km}
\bibitem{Cornwall:1974km}
  J.~M.~Cornwall, D.~N.~Levin and G.~Tiktopoulos,
   ``Derivation of Gauge Invariance from High-Energy Unitarity Bounds on the s
  Matrix,''
  Phys.\ Rev.\  D {\bf 10}, 1145 (1974)
  [Erratum-ibid.\  D {\bf 11}, 972 (1975)].
  %%CITATION = PHRVA,D10,1145;%%
  
  

  \bibitem{Peskin:1991sw}
  M.~E.~Peskin and T.~Takeuchi,
  ``Estimation of oblique electroweak corrections,''
  Phys.\ Rev.\  D {\bf 46}, 381 (1992).
  %%CITATION = PHRVA,D46,381;%%

%\cite{Barbieri:2004qk}
\bibitem{Barbieri:2004qk}
  R.~Barbieri, A.~Pomarol, R.~Rattazzi and A.~Strumia,
  ``Electroweak symmetry breaking after LEP-1 and LEP-2,''
  Nucl.\ Phys.\  B {\bf 703}, 127 (2004)
  [arXiv:hep-ph/0405040].
  %%CITATION = NUPHA,B703,127;%%



\bibitem{Flauger1}
%\cite{Flauger:2009ab}
%\bibitem{Flauger:2009ab}
  R.~Flauger, L.~McAllister, E.~Pajer, A.~Westphal and G.~Xu,
  ``Oscillations in the CMB from Axion Monodromy Inflation,''
  JCAP {\bf 1006} (2010) 009
  [arXiv:0907.2916 [hep-th]].
  %%CITATION = JCAPA,1006,009;%%


%\cite{Flauger:2010ja}
\bibitem{Flauger2}
  R.~Flauger, E.~Pajer,
  ``Resonant Non-Gaussianity,''
  JCAP {\bf 1101}, 017 (2011).
  [arXiv:1002.0833 [hep-th]]; see also
  %\cite{Hannestad:2009yx}
%\bibitem{Hannestad:2009yx}
  S.~Hannestad, T.~Haugbolle, P.~R.~Jarnhus and M.~S.~Sloth,
  ``Non-Gaussianity from Axion Monodromy Inflation,''
  JCAP {\bf 1006} (2010) 001
  [arXiv:0912.3527 [hep-ph]].
  %%CITATION = ARXIV:0912.3527;%%


\bibitem{Pajer}
  L.~Leblond and E.~Pajer,
  ``Resonant Trispectrum and a Dozen More Primordial N-point functions,''
  arXiv:1010.4565 [hep-th].
  %%CITATION = ARXIV:1010.4565;%%

\bibitem{Chen1}
  X.~Chen, R.~Easther and E.~A.~Lim,
  ``Large non-Gaussianities in single field inflation,''
  JCAP {\bf 0706}, 023 (2007)
  [arXiv:astro-ph/0611645].
  %%CITATION = JCAPA,0706,023;%%

%\cite{Chen:2008wn}
\bibitem{Chen2}
  X.~Chen, R.~Easther and E.~A.~Lim,
  ``Generation and Characterization of Large Non-Gaussianities in Single Field
  Inflation,''
  JCAP {\bf 0804}, 010 (2008)
  [arXiv:0801.3295 [astro-ph]].
  %%CITATION = JCAPA,0804,010;%%


\bibitem{Bean}
  R.~Bean, X.~Chen, G.~Hailu, S.~H.~Tye and J.~Xu,
  ``Duality Cascade in Brane Inflation,''
  JCAP {\bf 0803}, 026 (2008)
  [arXiv:0802.0491 [hep-th]].
  %%CITATION = JCAPA,0803,026;%%

\bibitem{Silverstein}
  E.~Silverstein and A.~Westphal,
  ``Monodromy in the CMB: Gravity Waves and String Inflation,''
  Phys.\ Rev.\  D {\bf 78}, 106003 (2008)
  [arXiv:0803.3085 [hep-th]].
  %%CITATION = PHRVA,D78,106003;%%

%\cite{Kobayashi:2010pz}
\bibitem{Kobayashi:2010pz}
  T.~Kobayashi, F.~Takahashi,
  ``Running Spectral Index from Inflation with Modulations,''
  JCAP {\bf 1101}, 026 (2011).
  [arXiv:1011.3988 [astro-ph.CO]].



%\cite{Komatsu:2001rj}
\bibitem{Komatsu}
  E.~Komatsu and D.~N.~Spergel,
  ``Acoustic signatures in the primary microwave background bispectrum,''
  Phys.\ Rev.\  D {\bf 63}, 063002 (2001)
  [arXiv:astro-ph/0005036].
  %%CITATION = PHRVA,D63,063002;%%

%\cite{Chen:2006nt}
\bibitem{Kachru}
  X.~Chen, M.~-x.~Huang, S.~Kachru, G.~Shiu,
  ``Observational signatures and non-Gaussianities of general single field inflation,''
  JCAP {\bf 0701}, 002 (2007).
  [hep-th/0605045].  



%\cite{Chen:2010bka}
\bibitem{Chen3}
  X.~Chen,
  ``Folded Resonant Non-Gaussianity in General Single Field Inflation,''
  JCAP {\bf 1012}, 003 (2010).
  [arXiv:1008.2485 [hep-th]].



%\cite{Maldacena:2002vr}
\bibitem{Maldacena:2002vr}
  J.~M.~Maldacena,
  ``Non-Gaussian features of primordial fluctuations in single field
  inflationary models,''
  JHEP {\bf 0305} (2003) 013
  [arXiv:astro-ph/0210603].
  %%CITATION = JHEPA,0305,013;%%


%\cite{Baumann:2010tm}
\bibitem{Baumann:2010tm}
  D.~Baumann, A.~Nicolis, L.~Senatore and M.~Zaldarriaga,
  ``Cosmological Non-Linearities as an Effective Fluid,''
  arXiv:1004.2488 [astro-ph.CO].
  %%CITATION = ARXIV:1004.2488;%%


\bibitem{Cosin}
  D.~Babich, M.~Creminelli, M.~Zaldarriaga,
  ``The Shape of Non-Gaussianities,''
  JCAP  {\bf0408} ,009 (2004).
  [astro-ph/0405356].



%---------------------




\end{thebibliography}
\end{document}